\documentclass[twocolumn,showpacs,preprintnumbers,amsmath,amssymb]{revtex4}
\usepackage{graphicx}
\usepackage{dcolumn}
\usepackage{bm}

\begin{document}

\title{
Charmonium properties in deconfinement phase in anisotropic lattice QCD
}
\author{H.~Iida$^1$\footnote{E-mail: iida@th.phys.titech.ac.jp}, 
T.~Doi$^2$, N.~Ishii$^3$, 
H.~Suganuma$^4$ and K.~Tsumura$^4$}

 \affiliation{$^1$Yukawa Institute for Theoretical Physics, Kyoto University, 
Kyoto 606-8502, Japan}
\affiliation{$^2$RIKEN BNL Research Center, 
Brookhaven National Laboratory, 
Upton, New York 11973, USA
}
\affiliation{$^3$
Department of Physics, The University of Tokyo, Tokyo 113-0033, Japan
}%
\affiliation{
$^4$Department of Physics, Kyoto University, 
Kitashirakawaoiwake, Sakyo, Kyoto 606-8502, Japan
}%

\date{\today}
\begin{abstract}
$J/\Psi$ and $\eta_c$ above the QCD critical temperature $T_c$ 
are studied in anisotropic quenched lattice QCD, 
considering whether the $c\bar c$ systems above 
$T_c$ are spatially compact (quasi-)bound states or scattering states. 
We adopt the standard Wilson gauge action and $O(a)$-improved Wilson 
quark action with renormalized anisotropy $a_s/a_t =4.0$ 
at $\beta=6.10$ on $16^3\times (14-26)$ lattices, 
which correspond to the spatial lattice volume 
$V\equiv L^3\simeq(1.55{\rm fm})^3$ and temperatures $T\simeq(1.11-2.07)T_c$. 
We investigate the $c\bar c$ system above $T_c$ 
from the temporal correlators 
with spatially-extended operators, where the overlap with the ground state 
is enhanced. 
To clarify whether compact charmonia survive in the deconfinement phase, 
we investigate spatial boundary-condition dependence of the energy of 
$c\bar c$ systems above $T_c$. 
In fact, for low-lying S-wave $c \bar c$ scattering states, 
it is expected that there appears a significant energy difference 
$\Delta E \equiv E{\rm (APBC)}-E{\rm (PBC)}
\simeq2\sqrt{m_c^2+3\pi^2/L^2}-2m_c$ ($m_c$: charm quark mass)  
between periodic and anti-periodic boundary conditions 
on the finite-volume lattice.
In contrast, for compact charmonia, there is no significant energy 
difference between periodic and anti-periodic boundary conditions. 
As a lattice QCD result, 
almost no spatial boundary-condition dependence is observed
for the energy of the $c\bar c$ system 
in $J/\Psi$ and $\eta_c$ channels  
for $T\simeq(1.11-2.07)T_c$.
This fact indicates that $J/\Psi$ and $\eta_c$ would survive as spatially 
compact $c\bar c$ (quasi-)bound states below $2T_c$.
We also investigate a $P$-wave channel at high temperature with maximally 
entropy method (MEM) and find no low-lying peak structure corresponding to 
$\chi_{c1}$ at $1.62T_c$. 
\end{abstract}

\maketitle

\section{Introduction}
\label{sec1}

 To complete the phase diagram of quantum chromodynamics (QCD) 
 is one of the most challenging attempts in particle physics. 
 The difficulty of QCD originates from the nonperturbative nature in the 
 low-energy region, where the running coupling constant becomes large. 
 As a consequence, color confinement and chiral symmetry breaking 
 occur as nonperturbative phenomena, and the vacuum becomes 
 the hadronic phase. 
 On the other hand, at high temperature or high density region, 
 color deconfinement and chiral symmetry 
 restoration are expected to be realized. 
 This phase is called the quark-gluon-plasma  (QGP) phase. 
Actually, lattice QCD simulations show color deconfinement 
\cite{CES83} and chiral symmetry restoration \cite{KSWGSSS83} 
above the QCD critical temperature $T_c$. 
The QGP phase transition 
is also investigated in various effective models \cite{HMHK86,MS86,
HK94,SST96,IOS05,Kapusta}. 

As an important signal of QGP creation, 
$J/\Psi$ suppression \cite{NA50,BRAHMS05,PHENIX05,PHOBOS05,STAR05} 
was theoretically proposed in the middle of 80's \cite{HMHK86,MS86}. 
The basic assumption of $J/\Psi$ suppression is that 
$J/\Psi$ disappears above $T_c$ due to vanishing of 
the confinement potential and appearance of the Debye screening effect, 
which are actually shown in lattice QCD simulations \cite{Rothe,K04}. 

Experimentally, QGP search is performed at CERN-SPS and RHIC 
with ultra-relativistic heavy-ion collisions 
\cite{BRAHMS05,PHENIX05,PHOBOS05,STAR05,NA50}. 
The experiment of NA-50 collaboration at CERN-SPS first reported the 
anomalous $J/\Psi$ suppression in Pb-Pb collision 
($158{\rm GeV}/c$ per nucleon)
 \cite{NA50}. 
Recently, the RHIC experiments show the various signals of QGP, 
e.g., $J/\Psi$ suppression \cite{HMHK86,MS86}, 
enhancement of strange particles \cite{RM82}, 
 jet quenching and high $p_T$ suppression \cite{Wang97,PHENIX03},
 elliptic flow $v_2$ \cite{PHENIX03} and so on, in 
the collisions of Au-Au ($200{\rm GeV}/c$ per nucleon). 

At first, QGP was naively speculated as simple quark-gluon gas. 
Nowadays, 
there are several indications that QGP is not simple 
perturbative quark-gluon gas in quenched lattice QCD. 
For example, 
it is pointed out
 that the spatial correlation in $\pi$ and $\sigma$ channels 
remains even in QGP phase in lattice QCD \cite{DK87}.
The other example is the relation between energy density and pressure, which 
does not satisfy the Stephan-Boltzmann relation even above $T_c$ \cite{Rothe}. 
The calculation of transport coefficients at finite temperature 
on quenched lattices 
also shows the strongly correlated gluon plasma \cite{NS05}. 
 These simulations indicate that some of the nonperturbative properties 
 may survive in QGP phase. 
Experimentally, 
in the reports of RHIC \cite{BRAHMS05,PHENIX05,PHOBOS05,STAR05}, 
 QGP seems to behave as perfect liquid, 
which strongly interacts, rather than dilute gas, from the comparison of 
the experiments with numerical simulations of hydrodynamics \cite{HN03}.
The strongly correlated deconfined phase is called strongly coupled 
QGP (sQGP) phase, and is investigated with much attention.

Very recently, some lattice QCD calculations indicate 
an interesting possibility that $J/\Psi$ and $\eta_c$ seem to 
survive even above $T_c$ \cite{UKMM01,UNM02,AH04,DKPW04}. 
 In Ref.~\cite{UKMM01}, the authors calculate correlators of 
 charmonia at finite temperature and find the strong spatial correlation 
 between $c$ and $\bar c$ even above $T_c$. 
In Refs.~\cite{UNM02,AH04,DKPW04}, the authors extract spectral functions 
of charmonia from temporal correlators 
at high temperature using the maximally entropy method (MEM). 
Although there are some quantitative differences, 
the peaks corresponding to $J/\Psi$ and $\eta_c$ seem to survive even 
above $T_c$ ($T_c< T< 2T_c$) in the $c\bar c$ spectral function. 

However, all of these calculations may suffer from a possible problem 
that the observed $c\bar c$ state on lattices 
 is not a nontrivial charmonium but a trivial $c\bar c$ scattering state, 
because it is difficult to distinguish these two states in lattice QCD. 
One of the reasons of the difficulty is that an narrow peak does not 
immediately indicate a spatially compact (quasi-)bound state.
In QGP phase, the potential between $q$ and $\bar q$ is considered to be 
the Yukawa potential due to the Debye screening~\cite{Rothe}. 
Therefore, the binding energy of $q$ and $\bar q$ above $T_c$ 
may be small. 
Then, the bound state of $q$ and $\bar q$ may not be spatially compact. 
In addition, MEM has a relatively large error, which sometimes 
leads to uncertainty for the structure of the spectral function. 

In this paper and our previous proceeding \cite{IIDS05}, using lattice QCD, 
we aim to clarify whether the $c\bar c$ system above $T_c$ 
is a spatially compact (quasi-)bound state or a scattering state, 
which is spatially spread. 
To distinguish these two states, we investigate spatial 
boundary-condition dependence 
of the energy of the $c\bar c$ system by 
comparing results on periodic and anti-periodic boundary conditions.
If the $c\bar c$ system is a scattering state,
there appears an energy difference $\Delta E$ between the 
 two boundary conditions as  
$\Delta E\simeq 2\sqrt{m_c^2+3\pi^2/L^2}-2m_c$ 
 with the charm quark mass $m_c$ on a finite-volume lattice with $L^3$ 
 (see Sec.~\ref{sec2}). 
 If the $c\bar c$ system is a spatially compact (quasi-)bound state, 
 the boundary-condition dependence is 
 expected to be small even in finite volume. 
In Ref.~\cite{IDIOOS05}, by changing the spatial periodicity 
of (anti-) quarks, the authors actually try to distinguish 
between a scattering state and a spatially compact resonance. 

In this study, we use anisotropic lattice QCD with anisotropy 
$a_s/a_t=4.0$. The reason why we use the anisotropic lattice is as follows. 
At finite temperature $T$, the temporal lattice size is restricted 
to $0\le t\le 1/T$. 
 Then, in calculating the temporal correlator $G(t)$ at high temperature, 
 we cannot take the sufficient number of points for $G(t)$. 
By using anisotropic lattice, 
more data points are available for $G(t)$. 
For the accurate measurement of $G(t)$, we use such a technical improvement 
of lattice QCD. 

 For further technical improvement, 
we use spatially-extended operators with hadron size 
in the actual lattice calculations at high temperature. 
We are interested in the low-lying spectrum in the 
$c\bar c$ systems at high temperature in this study, since 
 the ground-state component is desired to dominate in the 
range of $0\le t\le 1/T$. 
For this purpose, we use the spatially-extended operator 
to enhance the ground-state overlap. 

This paper is organized as follows. 
In Sec.~\ref{sec2}, we discuss the method to 
distinguish a spatially compact (quasi-)bound state from a 
scattering state by changing the spatial boundary condition for 
(anti-)quarks. 
In Sec.~\ref{sec3}
, we briefly explain anisotropic lattice QCD. 
 In Sec.~\ref{sec4}, 
we show the method to extract the energy 
of the ground state of $c\bar c$ systems from temporal correlators
 at finite temperature in lattice QCD. 
Section \ref{sec5} shows lattice QCD results of $J/\Psi$ and $\eta_c$ 
above $T_c$. Using the method discussed in Sec.~\ref{sec2}, we find the 
survival of $J/\Psi$ and $\eta_c$ as spatially compact (quasi-)bound states 
above $T_c (\sim 2T_c)$. 
In Sec.~\ref{sec6}, we perform the MEM analysis for the $c\bar c$ systems 
in $J/\Psi$, $\eta_c$ and $\chi_{c1}$ channels above $T_c$ using the lattice QCD 
data.  
 Section \ref{sec7} is devoted to conclusion and outlook. 

\section{Method to distinguish compact states from scattering states}
\label{sec2}

In this section, we explain the method to distinguish compact states from 
scattering states in term of their spatial extension. In Sec.~\ref{sec2-A}, 
we discuss and estimate the energy shift of a state due to the change of spatial boundary condition 
for the $c\bar c$ systems. 
In Sec.~\ref{sec2-B}, we discuss the correction from a short-range potential. 
As a result, we find that the correction is small compared with the energy shift in the 
$c \bar c$ scattering state above $T_c$. 
 
\subsection{Boundary condition dependence and energy shift for $c\bar c$ scattering states}
\label{sec2-A}

For the distinction between compact states and 
scattering states, we investigate the $c \bar c$ system 
on the periodic boundary condition (PBC) 
and on the anti-periodic boundary condition (APBC), respectively, 
and examine spatial boundary-condition dependence for the $c \bar c$ system. 
Here, in the PBC and the APBC cases, we impose 
periodic and anti-periodic boundary condition for (anti-)quarks 
on a finite-volume lattice, respectively. 

For a compact $c\bar c$ (quasi-)bound state, 
the wave function of the $c\bar c$ system is spatially localized 
and insensitive to spatial boundary conditions in lattice QCD 
as shown in Fig.~1(a), so that 
the charmonium behaves as a compact boson 
and its energy on APBC is almost the same as that on PBC \cite{IDIOOS05}. 

For a $c\bar c$ scattering state, the wave function is 
spatially spread and sensitive to spatial boundary conditions 
as shown in Fig.~1(b) and hence 
there emerges the energy difference between PBC and APBC 
due to the non-zero relative momentum of $c$ and $\bar c$ 
on APBC even in the lowest energy state.
On PBC, the momentum of a quark or an anti-quark is discretized as
\begin{eqnarray}
p_k=\frac{2n_k\pi}{L} \ \ (k=1,2,3, \ n_k\in {\bf Z})
\end{eqnarray}
on the finite lattice with the spatial volume $L^3$. 
Therefore, on PBC, the minimum momentum is $\vec p_{\rm min}=\vec 0$. 
On APBC, the (anti-)quark momentum is discretized as
\begin{eqnarray}
p_k=\frac{(2 n_k+1)\pi}{L} \ \ (k=1,2,3, \ n_k\in {\bf Z}).
\end{eqnarray}
In this case, the (anti-)quark momentum cannot take zero even in the 
lowest energy state. 
Then, the minimum (anti-)quark momentum is
\begin{eqnarray}
|\vec p_{\rm min}|=\frac{\sqrt{3}\pi}{L},
\end{eqnarray}
as is depicted in Fig.~2. 
Thus, there is an energy difference between PBC and APBC for 
the lowest $c\bar c$ scattering state. 
Neglecting the interaction between $c$ and $\bar c$, the energy difference 
is estimated as 
\begin{eqnarray}
\Delta E&\simeq& 2\sqrt{m_c^2+\vec p_{\rm min}^2}-2m_c\nonumber \\
&=&2\sqrt{m_c^2+3\pi^2/L^2}-2m_c, 
\end{eqnarray}
where $m_c$ is charm quark mass. The minimum momentum of a quark and 
an anti-quark 
in the case of scattering state is depicted in Fig.~2.

Table \ref{table1} summarizes the mass of the $c\bar c$ compact bound state 
(charmonia) 
and the energy 
of the $c\bar c$ scattering state both on PBC and APBC. 
In the case of the $c\bar c$ scattering state, 
there emerges the energy difference between PBC and APBC.

\begin{figure}
\rotatebox{-90}{
\includegraphics[width=5cm]{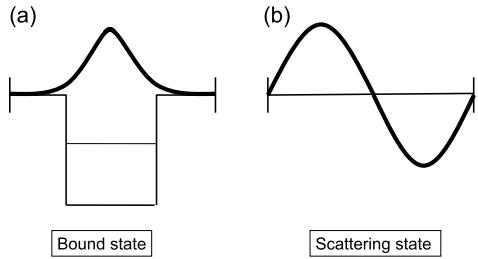}}
\caption{Schematic figures for 
boundary-condition dependence of a bound state (a) and 
a scattering state (b). 
The wave function of a bound state is spatially localized, and 
that of scattering state spatially spreads. Therefore, the bound state is 
insensitive to spatial boundary condition, while the scattering state is 
sensitive to spatial boundary condition in a finite-size box.}
\label{fig1}
\end{figure}
\begin{figure}
\rotatebox{-90}{
\includegraphics[width=3.5cm]{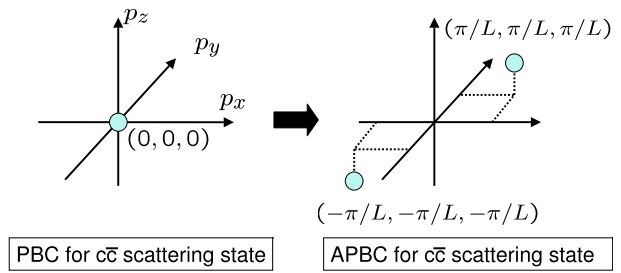}}
\caption{Pictorial expression of minimum momentum of the (anti-)quark 
on PBC and APBC in the case of $c\bar c$ scattering state 
in the center of mass flame. 
$c$ and $\bar c$ are in the finite-size with the spatial volume $L^3$. 
On PBC, both particles have zero lowest momentum in the lowest state. 
On the other hand, on APBC, quark and anti-quark have an opposite 
non-zero momentum.}
\label{fig1}
\end{figure}
\begin{figure}[h]
\rotatebox{-90}{\includegraphics[width=4cm]
{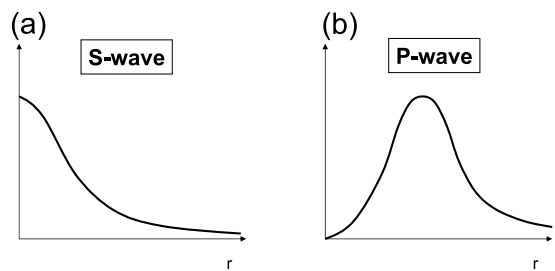}}
\caption{Schematic figure for 
radial wave functions of $c\bar c$ compact bound states for
the $S$-wave state (a) and the $P$-wave state (b). 
The $S$-wave state can be approximated with a Gaussian form 
$e^{-r^2/2\rho^2}$. 
In contrast, the wave-function of $P$-wave state 
should be zero at $r=0$ and spatially spreads due to the 
centrifugal potential.}
\label{RWF}
\end{figure}
\begin{table*}
\begin{center}
\caption{
Summary of the boundary-condition dependence 
of the compact charmonia and the $c\bar c$ scattering 
state. The energy difference $\Delta E$ is calculated on the 
spatial lattice size $L=1.55{\rm fm}$ and the charm quark mass 
$m_c\simeq 1.3{\rm GeV}$. 
}
\label{table1}
\begin{tabular}{lll}
\hline
\hline
& \ \ charmonia \ \ & \ \ $c\bar c$ scattering state \ \ \\
\hline
mass (energy) on PBC \ \ & \ \ bound state mass $M$ \ \ & \ \
 $\simeq 2m_c$ \ \ \\
mass (energy) on APBC \ \ & \ \ bound state mass $M$ \ \ & \ \ 
$\simeq 2\sqrt{m_c^2+\vec p_{\rm min}^2}$ \ \ \\
$\Delta E\equiv E({\rm APBC})-E({\rm PBC})$& $ \ \ \simeq 0 \ \ $& \ \ 
$\simeq 2\sqrt{m_c^2+\vec p_{\rm min}^2}-2m_c 
\simeq 0.35{\rm GeV}$ \ \ \\
\hline
\hline
\end{tabular}
\end{center}
\end{table*}

\begin{table*}
\begin{center}
\caption{The lowest energy of the $c\bar c$ scattering state 
above $T_c$ on PBC and APBC. $E_0^{\rm free}$ and $E^{\rm Y}_0$ denote the 
lowest energy in the no interaction (free) and the Yukawa-potential cases, respectively 
(the origin of the energy is shifted by $2m_c$ in this non-relativistic estimation). 
The energy difference from the free case, $E_0^{\rm Y}-E_0^{\rm free}$, is also shown.}
\label{scattering}
\begin{tabular}{cccc}
\hline\hline
&$E_0^{\rm free}$ \ & \ $E_0^{\rm Y}$ \ & \ $E_0^{\rm Y}-E_0^{\rm free}$\\
\hline
PBC \ \ &$0{\rm MeV}$ \ \ &$-7.8{\rm MeV}$&$-7.8{\rm MeV}$\\
APBC \ \ & $318.4{\rm MeV}$ \ \ &$304.7{\rm MeV}$&$-13.7{\rm MeV}$\\
\hline\hline
\end{tabular}
\end{center}
\end{table*}

\begin{table*}
\begin{center}
\caption{Lattice parameters and related quantities 
in our anisotropic lattice QCD calculation.}
\label{table3}
\begin{tabular}{ccccccccc}
\hline
\hline
$\beta$  & lattice size & $a_s^{-1}$ & $a_t^{-1}$ 
&$\gamma_G$& $u_s$ & $u_t$ & $\gamma_F$ & $\kappa$
\\
\hline
6.10 & $16^3 \times(14-26)$& 2.03GeV & 8.12GeV&3.2103&0.8059 
&0.9901 &4.0 & 0.112\\
\hline
\hline
\end{tabular}
\end{center}
\end{table*}


 Such a method, to distinguish a compact (quasi-) bound state from 
a scattering state by changing boundary conditions, is actually used in 
Ref.~\cite{IDIOOS05}.
 This method is essentially 
based on the finiteness of lattice volume. 
Note that 
the finite volume is used in several studies for the analyses 
of a scattering state and/or a compact bound state in lattice QCD
\cite{L91, MLABCDDHLTZ04, TUOK05}. 


\subsection{Correction from a short-range potential}
\label{sec2-B}

Here, we discuss the possible correction to the energy difference of a scattering state 
between PBC and APBC from a short-range potential. In the previous subsection, 
we neglect the interaction between $c$ and $\bar c$  
and estimate the energy difference between PBC and APBC for the $c \bar c$ scattering state. 
However, in the actual situation above $T_c$, 
a quark and anti-quark interact with each other 
by the Yukawa potential
\begin{align}
V(r)=-A\frac{ e^{-m(T) r}}{r},
\label{Yukawa}
\end{align}
where $m(T)$ is the temperature-dependent Debye screening mass \cite{Rothe,G90}. 
The prefactor $A$ corresponds to the Coulomb coefficient at $T=0$ and is estimated as $A\simeq$ 0.28 
\cite{TMNS01}.
Therefore, the simple estimation of the energy shift $\Delta E$ 
obtained in Sec.~\ref{sec2-A} may be corrected in the presence of the Yukawa 
potential. In the following, considering the Yukawa potential $V(r)$, 
we estimate the energy shift $\Delta E$ of the $c\bar c$ scattering 
state in the non-relativistic quantum mechanics.

Consider two particles, $c$ and $\bar c$, in the finite box with 
$x,y,z\in [0,L]$, where 
the boundary condition is periodic or anti-periodic. 
$c$ and $\bar c$ interact each other with the 
Yukawa potential in Eq.~(\ref{Yukawa}). 
We estimate the lowest energy of the $c \bar c$ scattering state 
in PBC and APBC cases, respectively. 

We use the variational method for the charmonium wave function $\psi({\vec r})$ (${\vec r}$: the relative coordinate) 
in this estimate. We prepare appropriate basis $\phi_i({\vec r})$ and expand the wave function $\psi({\vec r})$ as 
\begin{align}
\psi({\vec r})=\sum_i C_i \phi_i({\vec r}),
\end{align}
where the coefficients $\{C_i\}$ satisfy $\sum_i|C_i|^2=1$, for the orthonormal basis 
$\phi_i({\vec r})$. 
The hamiltonian in the $c\bar c$ system is given by
\begin{align}
&\hat H=-\frac{1}{2\mu}\left(\frac{\partial^2}{\partial x^2}
+\frac{\partial^2}{\partial y^2}
+\frac{\partial^2}{\partial z^2}\right)
+V(r), 
\label{Energy}
\end{align}
where $\mu=m_c/2$ is the reduced mass of the $c\bar c$ system. 
The energy of the system is expressed by 
\begin{align}
&E\equiv \frac{\int_V d^3{r} \ \psi^\dagger ({\vec r}) \hat H \psi({\vec r})}
{\int_V d^3{r} \ \psi^\dagger({\vec r})\psi({\vec r})}.
\label{energy}
\end{align}
Differentiating Eq.~({\ref{energy}}) by $C_i$ and imposing   
the stationary condition for $E$, i.e., $\frac{\partial E}{\partial C_i}=0$, 
we get the equation for $E$ and $C_i$ as
\begin{align}
&\sum_jC_j \left\{\int_V d^3{r} \ \phi^\dagger_i ({\vec r})\hat H\phi_j({\vec r})
-E \int_V d^3{r} \ \phi_i^\dagger ({\vec r}) \phi_j({\vec r})\right\}=0.
\end{align}
From the condition that $\{C_i\}$ have non-trivial solutions, $E$ is determined as 
\begin{align}
{\rm det}(H_{ij}-E S_{ij})=0,
\label{variation}
\end{align}
where $H_{ij}\equiv \int_V d^3{r} \ \phi^\dagger_i({\vec r})\hat H\phi_j({\vec r})$ and 
$S_{ij}\equiv \int_V d^3{r} \ \phi^\dagger_i({\vec r})\phi_j({\vec r})$. 
Solving Eq.~(\ref{variation}), the energy of the $c\bar c$ scattering state is obtained.

We take the $i$-th basis as follows:
\begin{align}
\phi_i(\vec {r})&=
\sqrt{\frac{(2-\delta_{n_x0})(2-\delta_{n_y0})(2-\delta_{n_z0})}{L^3}} \nonumber\\
&\times \cos \left(\frac{2\pi n_x}{L} x\right) 
\cos\left(\frac{2\pi n_y}{L}
 y\right) \cos \left(\frac{2\pi n_z}{L} z\right), 
\end{align}  
where the normalization factor is set so as to satisfy the orthonormal condition, 
$\int d^3{r} \ \phi_i^\dagger \phi_j=\delta_{ij}$.
In the PBC case, we choose the basis as 
\begin{align}
&\phi_0: {\vec n}=(0,0,0)\nonumber\\
&\phi_1,\phi_2,\phi_3: {\vec n}=(1,0,0), (0,1,0), (0,0,1)\nonumber\\
&\phi_4,\phi_5,\phi_6: {\vec n}=(0,1,1), (1,0,1), (1,1,0)\nonumber\\
&\cdots .\nonumber
\end{align}
Note that $\phi_1$, $\phi_2$ and $\phi_3$ degenerate and so on. 
In the APBC case, we choose the basis as
\begin{align}
&\phi_0: {\vec n}=\left(\frac{1}{2},\frac{1}{2},\frac{1}{2}\right)\nonumber\\
&\phi_1,\phi_2,\phi_3: {\vec n}=\left(\frac{3}{2},\frac{1}{2},\frac{1}{2}\right),
\left(\frac{1}{2},\frac{3}{2},\frac{1}{2}\right),\left(\frac{1}{2},\frac{1}{2},\frac{3}{2}\right)\nonumber\\
&\phi_4,\phi_5,\phi_6: {\vec n}=\left(\frac{1}{2},\frac{3}{2},\frac{3}{2}\right),
\left(\frac{3}{2},\frac{1}{2},\frac{3}{2}\right),\left(\frac{3}{2},\frac{3}{2},\frac{1}{2}\right)\nonumber\\
&\cdots .\nonumber
\end{align}
In the following, we consider the $S$-wave case. Due to the spherical symmetry of $S$-wave, 
the wave function $\psi$ is invariant under the replacement of $x\leftrightarrow y, 
y\leftrightarrow z, z\leftrightarrow x$. Therefore, for instance,  
the coefficients of $\phi_1, \phi_2$ and $\phi_3$ coincide, and one finds $C_1=C_2=C_3$, 
$C_4=C_5=C_6$ and so on, both in PBC and APBC cases. 
The convergence for the number of basis, $N_{\rm basis}$, is checked and 
the difference of the energy of lowest state is found to be less than 1{\rm MeV} between 
$N_{\rm basis}=7$ and $13$.

We use $A=0.28$ as the Coulomb coefficient in $T=0$ case. 
We adopt $m(T)=gT$ as the result of the lowest-order perturbative QCD calculation at finite temperature 
\cite{Hatsuda}. 
For the QCD coupling constant $g$, we use $g=1$ as a typical value in the infrared region, 
which corresponds to the scale in the typical lattice QCD simulations with $\beta\equiv \frac{2N_c}{g^2}\sim 6$. 
In this calculation, we consider the case at $T=300{\rm MeV}$, which is slightly above $T_c$ in quenched QCD. 
The mass of charm quark is set by $m_c=1.5{\rm GeV}$. 

Table~{\ref{scattering}} 
is the summary of the results. $E_0^{\rm free}$ denotes the energy of 
the lowest state in free case. 
$E_0^{\rm Y}$ denotes the lowest-state energy in the 
Yukawa potential. Note that the energy difference $|E_0^{\rm Y}-E_0^{\rm free}|$ is rather small 
as $7.8{\rm MeV}$ in PBC case and $13.7{\rm MeV}$ in APBC case, i.e.,
 about $10{\rm MeV}$. 
(Even in the extreme case of the Coulomb potential, i.e., $m(T)=0$, 
where the potential effect is clearly overestimated, 
the correction is found to be 50{\rm MeV} at most.)
 
Thus, the correction from the short-range potential about $10{\rm MeV}$ 
is enough small compared with the energy shift $\Delta E\simeq 350{\rm MeV}$ between PBC and APBC 
shown in Table~\ref{table1}. 
Therefore, as far as we consider the $c\bar c$ scattering 
state above $T_c$, short-range interaction between $c$ and $\bar c$ 
can be neglected. 

Here, we note that the parameter set used in this calculation gives the upper limit of the 
correction from the short-range Yukawa interaction. 
As the temperature increases, the Debye screening mass $m(T)$ becomes larger and the ``Coulomb coefficient" $A(T)$
 becomes slightly smaller. 
These tendencies make the interaction smaller at high temperature. 
Thus, using $A(T)$ at $T=0$ and setting the lowest temperature $T\simeq T_c$, which realize the QGP phase, 
we get the upper limit of the correction from the short-range interaction. 
In addition, the use of the perturbative estimation $m(T)\simeq gT$ also leads an 
``overestimation" for the effect of the Yukawa potential. For, the lattice QCD calculation \cite{G90} 
shows that the actual Debye screening is estimated as $m(T)=(2-3)gT$ near $T_c$ and the $q\bar q$ potential 
seems to be more screened the perturbative result. 
In fact, the obtained correction about $10{\rm MeV}$ is reduced in the realistic case.
Through the above considerations, the correction obtained in the calculation can be 
regarded as the  
upper limit and the actual value of the correction should be smaller than the limit about $10{\rm MeV}$.


\begin{figure*}[t]
\rotatebox{-90}{\includegraphics[width=5.3cm]
{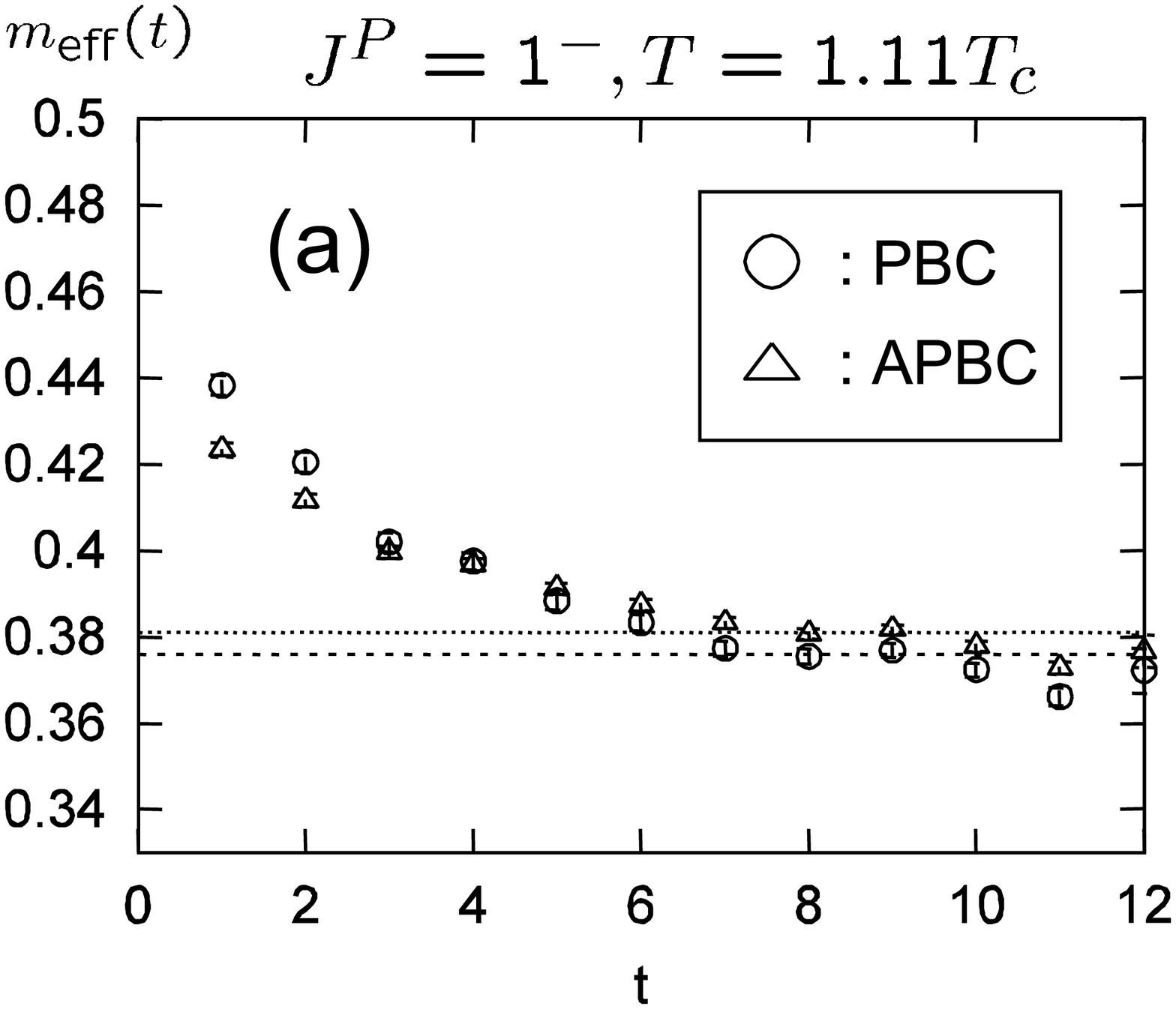}}
\rotatebox{-90}{\includegraphics[width=5.2cm]
{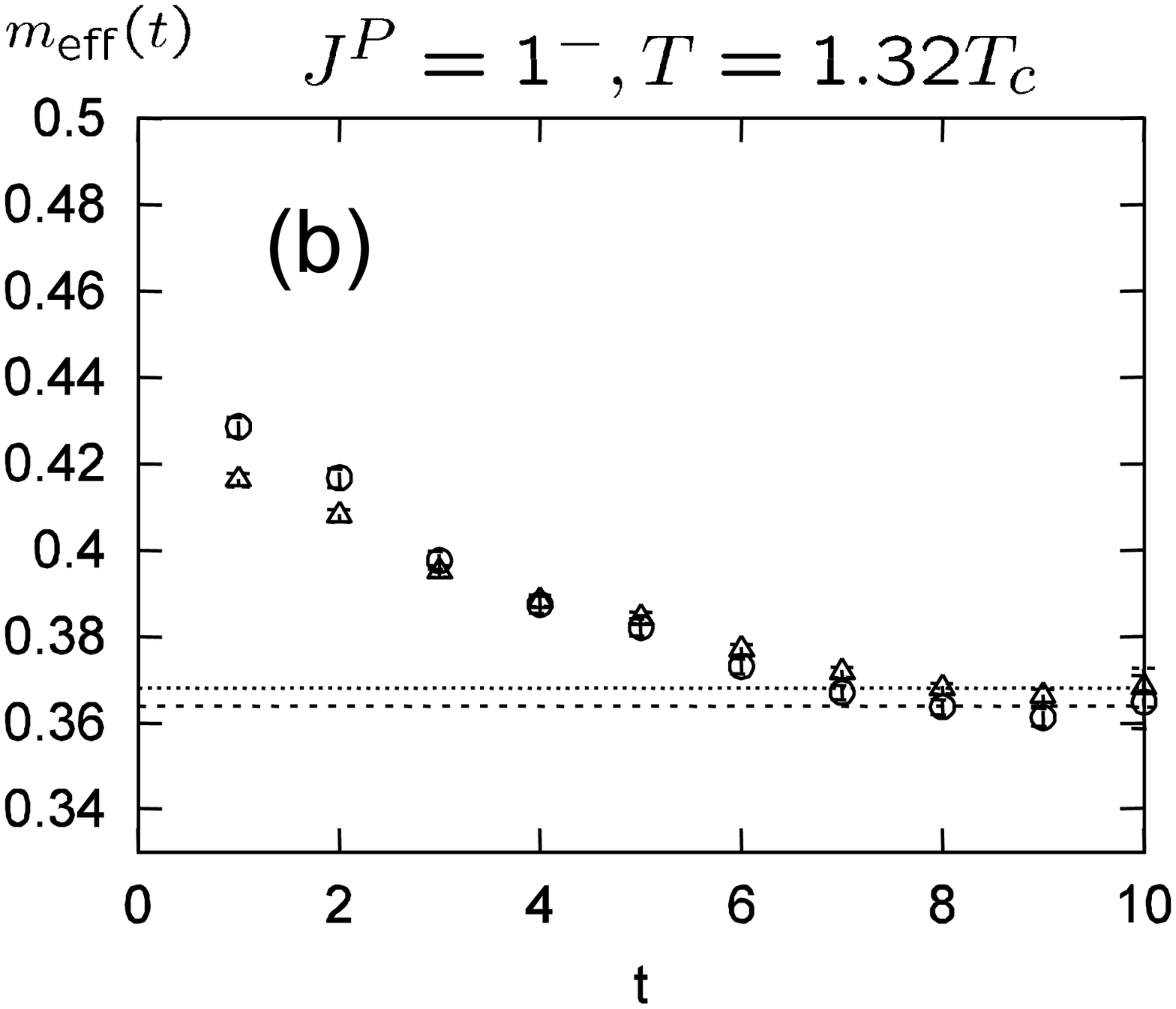}}
\rotatebox{-90}{\includegraphics[width=5.2cm]
{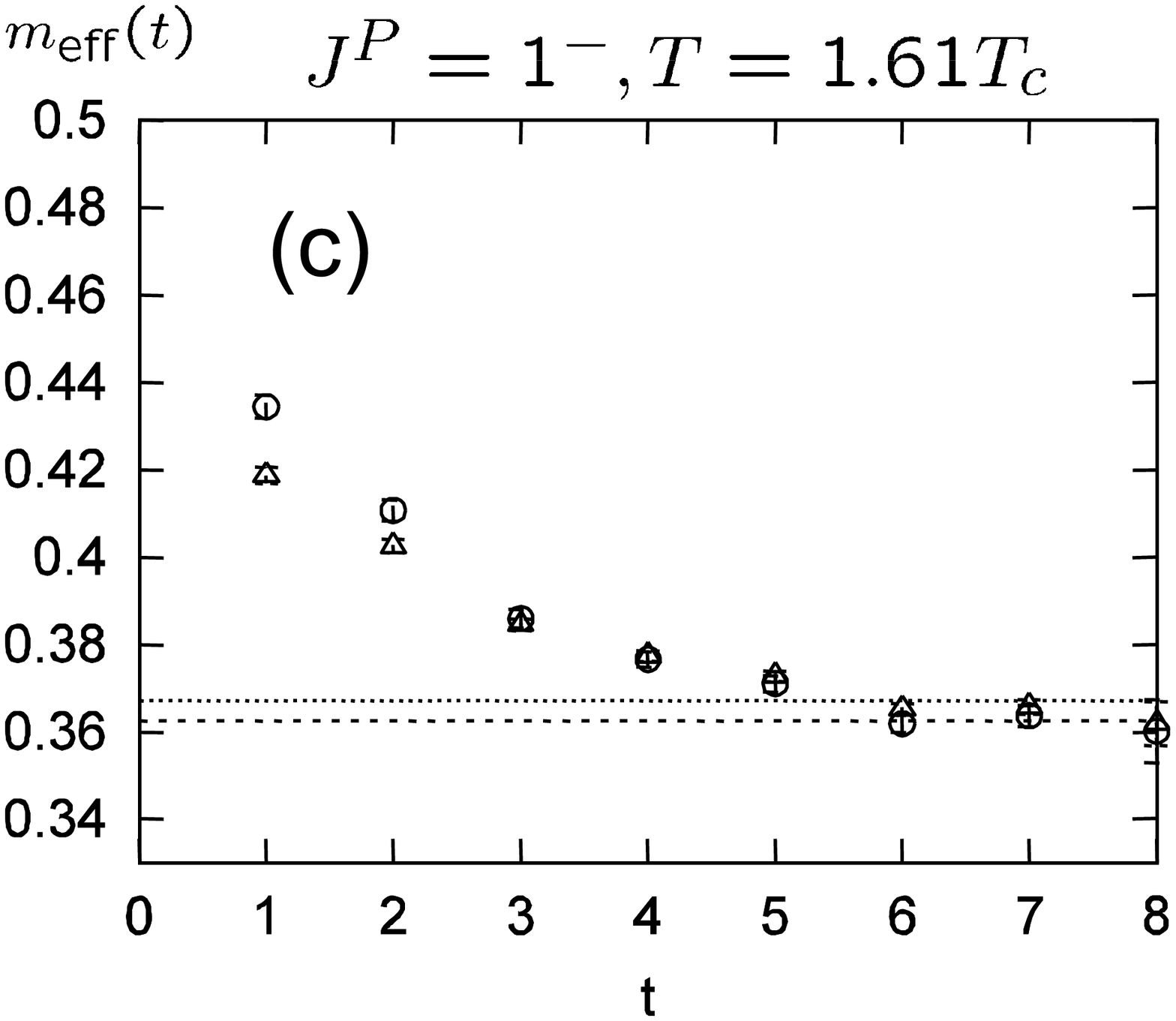}}
\rotatebox{-90}{\includegraphics[width=5.2cm]
{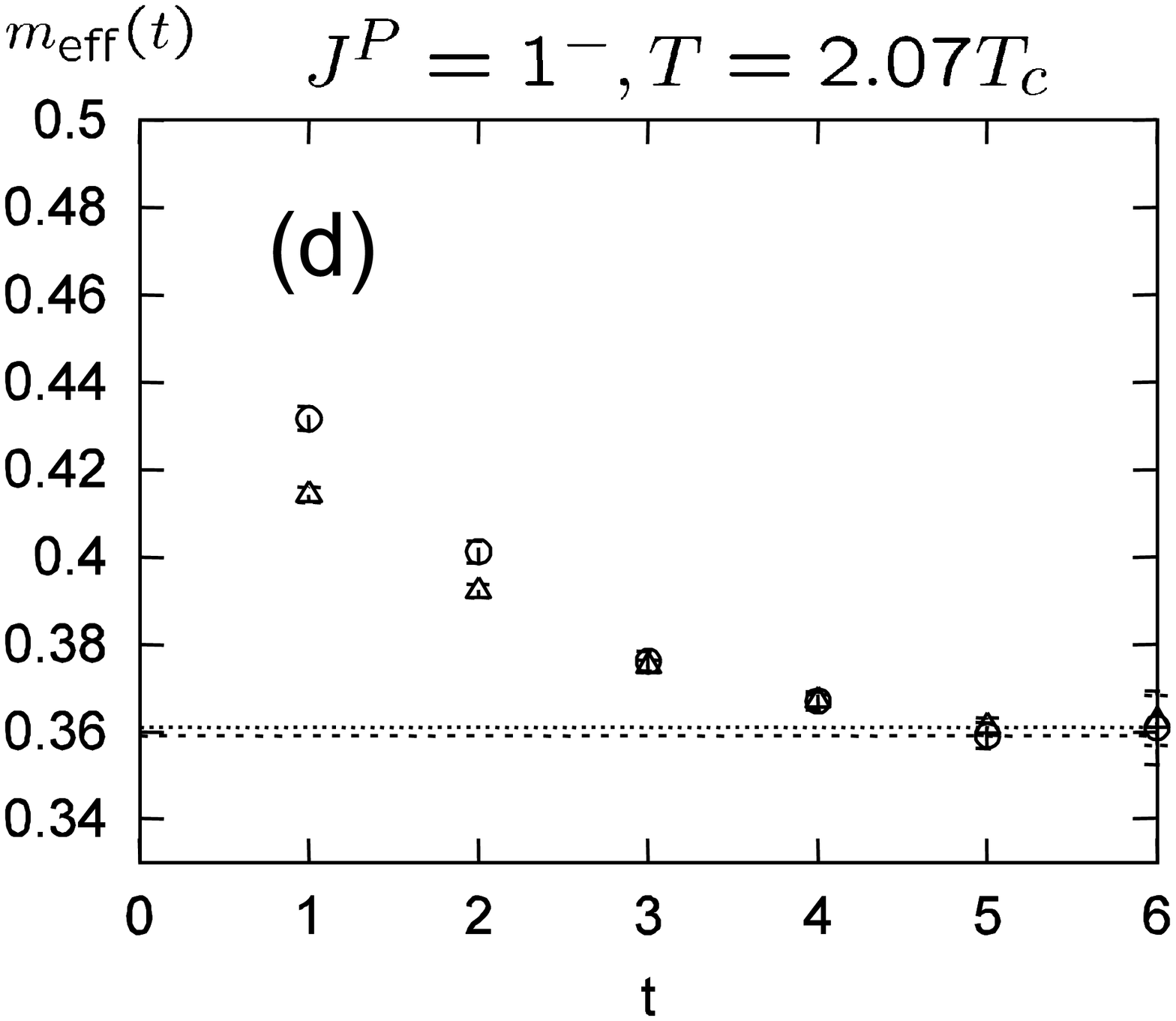}}
\caption{Effective masses of $J/\Psi$ at $1.11T_c$ (a), $1.32T_c$ (b), 
$1.61T_c$ (c) and $2.07T_c$ (d) in the lattice unit 
with $a_t=(8.12{\rm GeV})^{-1}$.
 The circles and the triangles denote the results on PBC and APBC, 
 respectively.
 The dashed and dotted lines denote $E$(PBC) and $E$(APBC) 
 obtained from the best-fit analysis, respectively.
}
\label{fig2}
\end{figure*}

\begin{table*}
\begin{center}
\caption{The energy of the $c\bar c$ system 
in the $J/\Psi$ channel ($J^P=1^{-}$)
on PBC and APBC at $\beta=6.10$ and $\rho=0.2{\rm fm}$ at each temperature. 
The statistical errors are smaller than 0.01GeV.
We list also uncorrelated $\chi^2/N_{\rm DF}$ and 
$\Delta E\equiv E{\rm (APBC)}-E{\rm (PBC)}$.}
\label{table5}
\begin{tabular}{lllll}
\hline
\hline
temperature & fit range & $E$(PBC) [$\chi^2/N_{\rm DF}$] 
& $E$(APBC) [$\chi^2/N_{\rm DF}$]
 & $\Delta E$\\
\hline
$1.11T_c$ & 7--11  &3.05{\rm GeV} [0.14] & 3.09{\rm GeV} [0.61] &0.04{\rm GeV}
\\
$1.32T_c$ & 8--11 &2.95{\rm GeV} [0.34] & 2.98{\rm GeV} [0.33] &0.03{\rm GeV}\\
$1.61T_c$ & 6--9  &2.94{\rm GeV} [0.10] & 2.98{\rm GeV} [0.22]
 &0.04{\rm GeV}\\
$2.07T_c$ & 5--7  &2.91{\rm GeV} [0.03] & 2.93{\rm GeV} [0.04]
 &0.02{\rm GeV}\\
\hline
\hline
\end{tabular}
\end{center}
\end{table*}
\begin{figure}[t]
\rotatebox{-90}{\includegraphics[width=5.2cm]
{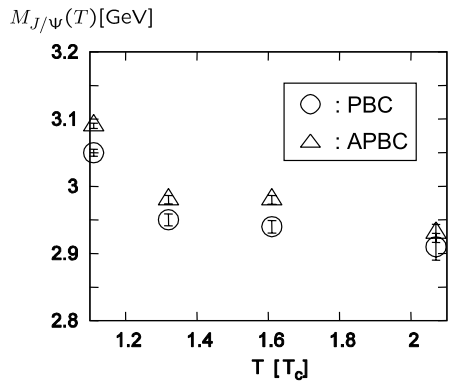}}

\caption{Temperature dependence of the pole mass (energy) $M_{J/\Psi}(T)$ 
 of $J/\Psi$ for $(1.11-2.07)T_c$ on PBC (circles) and APBC (triangles). 
The circles and the triangles correspond to $E$(PBC) and $E$(APBC) 
of the $c\bar c$ system, respectively. 
 The energy difference $\Delta E\equiv E({\rm APBC})-E({\rm PBC})\simeq 
 (0.02-0.04){\rm GeV}$ between 
PBC and APBC are considerably smaller than that of $c\bar c$ 
scattering state, $\Delta E\simeq 0.35{\rm GeV}$.}
\label{fig6}
\end{figure}

\begin{figure*}[hbt]
\rotatebox{-90}{\includegraphics[width=5.2cm]
{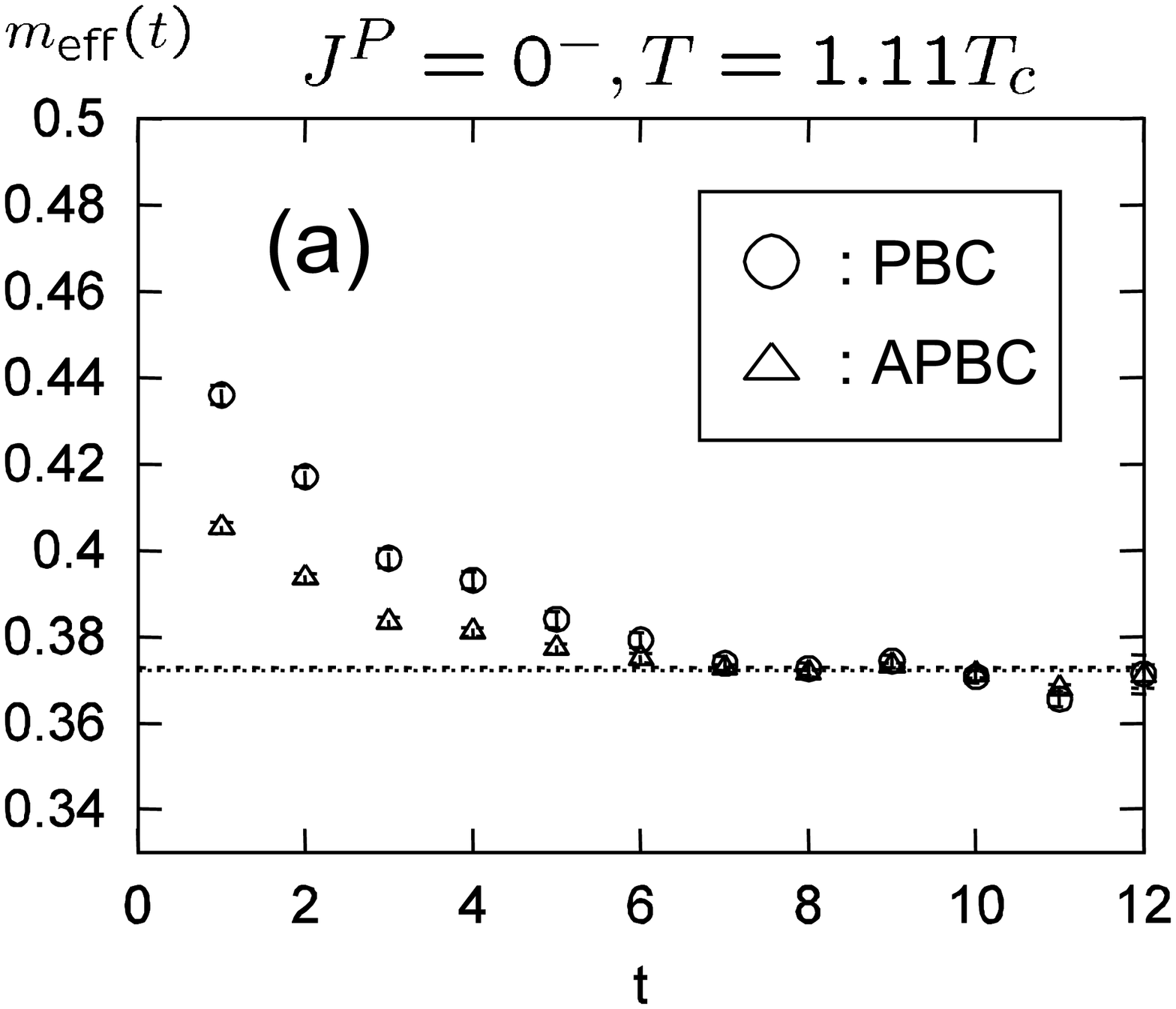}}
\rotatebox{-90}{\includegraphics[width=5.2cm]
{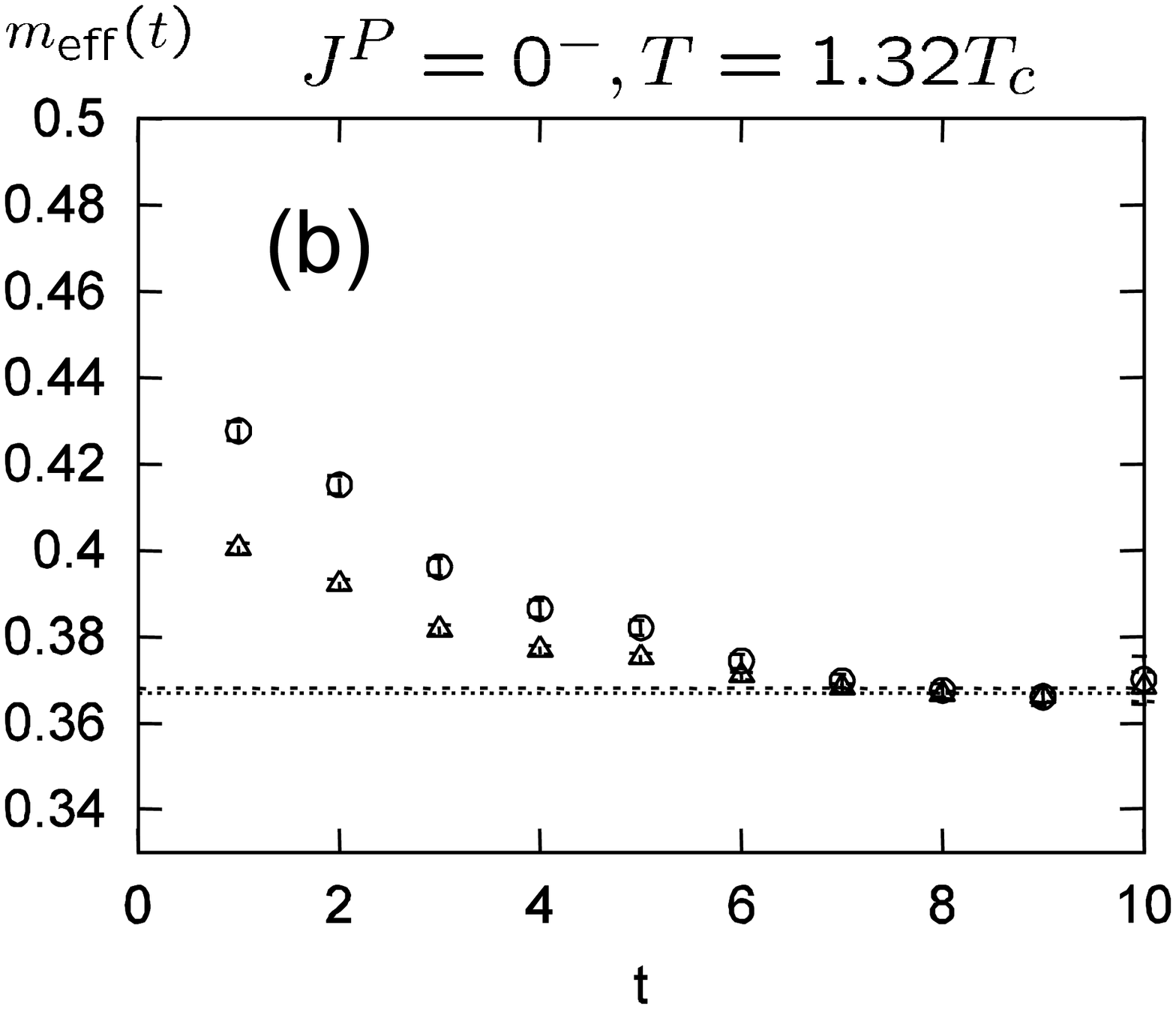}}
\rotatebox{-90}{\includegraphics[width=5.2cm]
{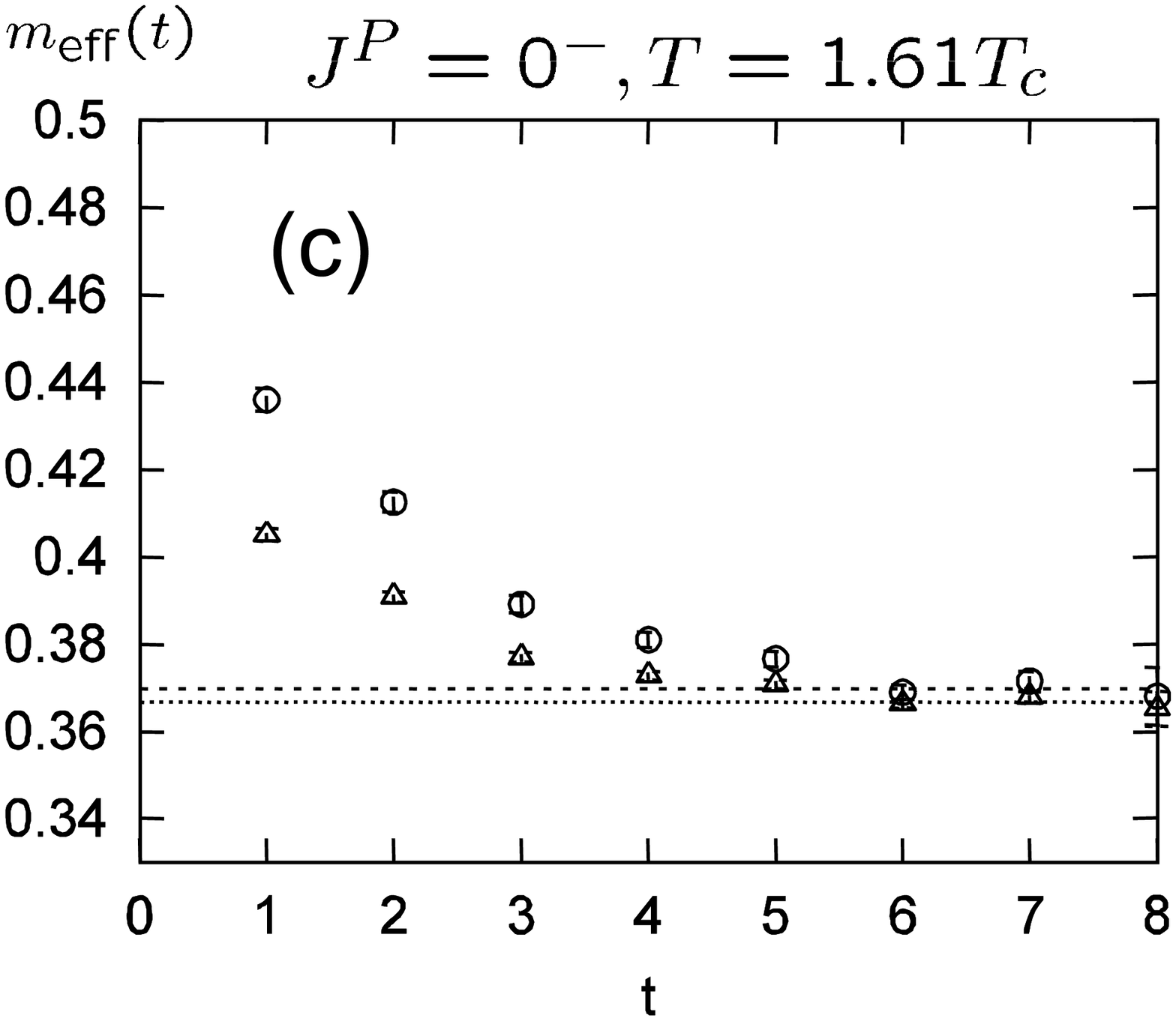}}
\rotatebox{-90}{\includegraphics[width=5.2cm]
{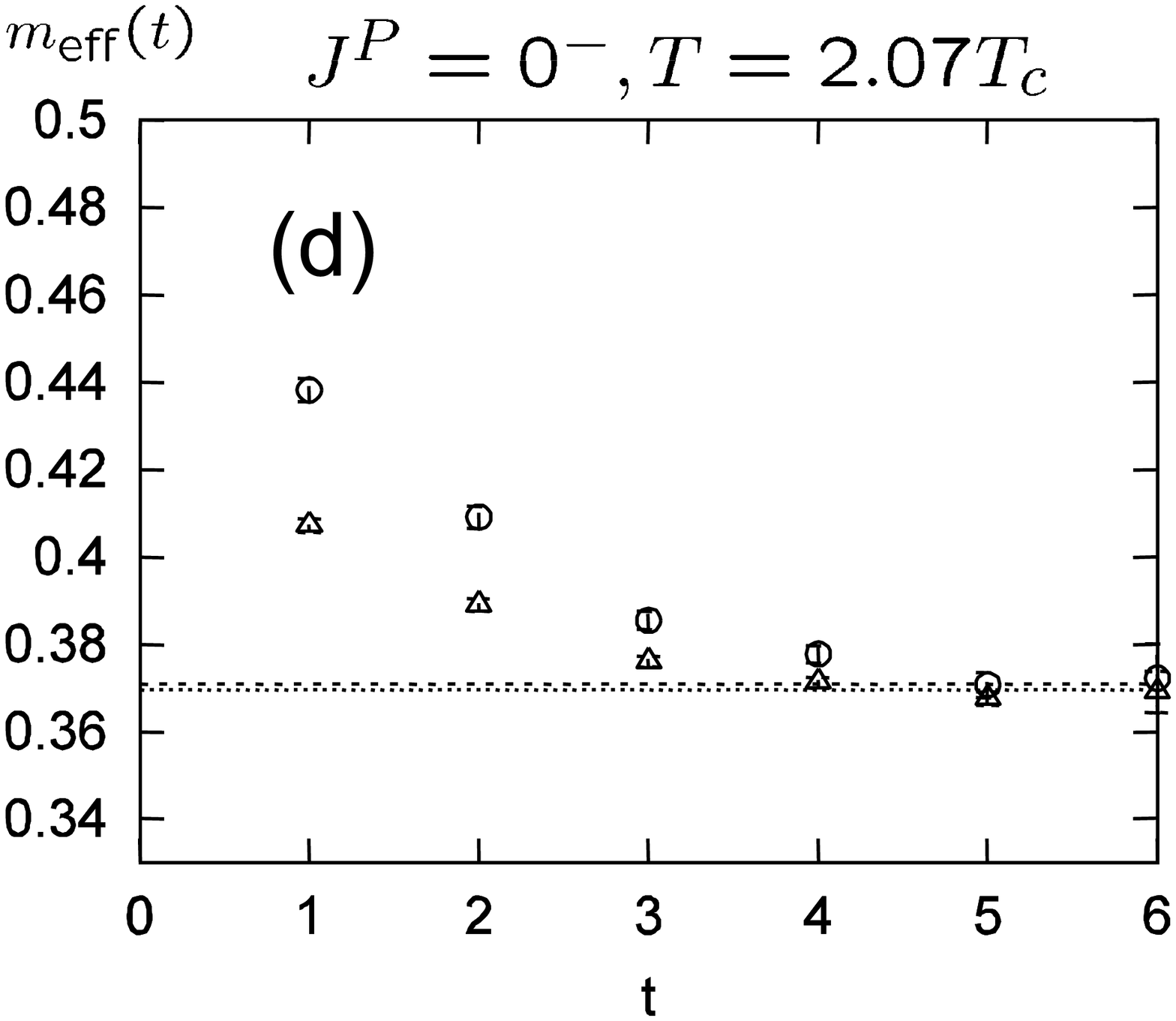}}
\caption{Effective masses of $\eta_c$ at $1.11T_c$ (a), $1.32T_c$ (b), 
$1.61T_c$ (c) and $2.07T_c$ (d) in the lattice unit with $a_t\simeq 
(8.12{\rm GeV})^{-1}$.
 The circles and the triangles denote the results on PBC and APBC, 
 respectively.
 The dashed and the dotted lines denote $E$(PBC) and $E$(APBC) obtained 
 from the best-fit analysis, respectively.
}
\label{fig7}
\end{figure*}

\begin{table*}
\caption{The energy of the $c\bar c$ system in the 
$\eta_c$ channel ($J^P=0^{-}$) 
on PBC and APBC at $\beta=6.10$ and $\rho=0.2{\rm fm}$ at each temperature.
 The statistical errors are smaller than 0.01GeV.
We list also uncorrelated $\chi^2/N_{\rm DF}$ and 
$\Delta E\equiv E{\rm (APBC)}-E{\rm (PBC)}$.}
\label{table6}
\begin{center}
\begin{tabular}{lllll}
\hline
\hline
temperature & fit range & $E$(PBC) [$\chi^2/N_{\rm DF}$] & $E$(APBC)  
[$\chi^2/N_{\rm DF}$] & $\Delta E$\\
\hline
$1.11T_c$ &7--11  &3.03{\rm GeV} [0.04] & 3.02{\rm GeV} 
[0.17]  &-0.01{\rm GeV}\\
$1.32T_c$ & 7--11 &2.99{\rm GeV} [0.78] & 2.98{\rm GeV} 
[0.82] &-0.01{\rm GeV}\\
$1.61T_c$ & 6--9  &3.00{\rm GeV} [0.31] & 2.97{\rm GeV} 
[0.38] &-0.03{\rm GeV}\\
$2.07T_c$ & 5--7  &3.01{\rm GeV} [0.03] & 3.00{\rm GeV} 
[0.07] &-0.01{\rm GeV}\\
\hline
\hline
\end{tabular}
\end{center}
\end{table*}

\section{Anisotropic lattice QCD}
\label{sec3}

In this paper, we adopt anisotropic lattice QCD 
for the study of high-temperature QCD, as was mentioned in Sec.~\ref{sec1}. 
We explain anisotropic lattice QCD in the following. 

For the gauge field, we adopt the standard plaquette action 
on an anisotropic lattice 
as \cite{IDIOOS05,MOU01}
\begin{eqnarray}
S_G&=&\frac{\beta}{N_c}\frac{1}{\gamma_G}
\sum_{s,i<j\leq 3} {\rm Re}{\rm Tr}\{1-P_{ij}(s)\}\nonumber \\
&+& \frac{\beta}{N_c}\gamma_{G}\sum_{s,i\leq 3}{\rm Re}{\rm Tr}\{1-P_{i4}(s)\},
\end{eqnarray}
where $P_{\mu\nu}$ denotes the plaquette operator.
In the simulation, we take $\beta \equiv 2N_c/g^2=6.10$ and  
the bare anisotropy $\gamma_G=3.2103$, 
which lead to the renormalized anisotropy as $a_s/a_t=4.0$.
The scale is set by the Sommer scale as $r_0^{-1}=395{\rm MeV}$. 
Then, the spatial and the temporal lattice spacing are evaluated as 
$a_s^{-1}\simeq 2.03{\rm GeV}$ (i.e., $a_s\simeq$ 0.097fm),
and $a_t^{-1}\simeq 8.12{\rm GeV}$ (i.e., $a_t\simeq$ 0.024fm), respectively.
The adopted lattice sizes are $16^3\times (14-26)$, 
which correspond to the spatial lattice size as $L \simeq 1.55{\rm fm}$ 
and the temperature as $(1.11-2.07)T_c$. 
Here, the critical temperature is estimated as $T_c=(260-280){\rm MeV}$ 
at the quenched level \cite{ISM02}. 
We use 999 gauge configurations, which are picked up every 
500 sweeps after the thermalization of 20,000 sweeps. 
We adopt the jackknife error estimate for lattice data. 
 
For quarks, we use $O(a)$-improved Wilson (clover) action 
on the anisotropic lattice as \cite{IDIOOS05,MOU01}
\begin{eqnarray}
&&S_F \equiv \sum_{x,y}\bar \psi(x)K(x,y)\psi(y), \nonumber \\
&&K(x,y) \equiv \nonumber \\
&&\delta_{x,y}
\kappa_t\{(1-\gamma_4)U_4(x)\delta_{x+\hat 4,y}
+(1+\gamma_4)U_4^\dagger(x-\hat 4)\delta_{x-\hat 4,y}\}\nonumber \\
&&- \kappa_s \sum_i \{ (r-\gamma_i)U_i(x)\delta_{x+\hat i,y}\}
+(r+\gamma_i)U_i^\dagger(x-\hat i)\delta_{x-\hat i,y}\} \nonumber \\
&&-\kappa_s c_E \sum_i \sigma_{i4}F_{i4}\delta_{x,y}-r\kappa_s c_B\sum_{i<j}
\sigma_{ij}F_{ij}\delta_{x,y},
\end{eqnarray}
which is anisotropic version of the Fermilab action \cite{EKM97}.
$\kappa_s$ and $\kappa_t$ denote the spatial and the temporal hopping 
parameters, respectively, and $r$ the Wilson parameter. 
$c_E$ and $c_B$ are the clover coefficients.
The tadpole improvement is done by the replacement of $U_i(x)\rightarrow U_i(x)
/u_s$, $U_4(x)\rightarrow U_4(x)/u_t$, where $u_s$ and $u_t$ are 
the mean-field values of the spatial and the temporal 
link variables, respectively. 
The parameters $\kappa_s, \kappa_t, r, c_E, c_B$ are to be tuned 
so as to keep the Lorentz symmetry up to $O(a^2)$.
At the tadpole-improved tree-level, this requirement leads to 
$r=a_t/a_s$, $c_E=1/(u_su_t^2)$, $c_B=1/u_s^3$ and 
the tuned fermionic anisotropy 
$\gamma_F\equiv (u_t \kappa_t)/(u_s \kappa_s)=a_s/a_t$.
For the charm quark, we take $\kappa=0.112$ with 
$1/\kappa \equiv 1/(u_s \kappa_s)-2(\gamma_F+3r-4)$, 
which corresponds to the hopping parameter in the isotropic lattice.
The bare quark mass $m_0$ in spatial lattice unit is expressed 
as $m_0=\frac12(\frac{1}{\kappa}-8)$.
We summarize the lattice parameters and related quantities in Table 
\ref{table3}.
With the present lattice QCD setup, the masses of $J/\Psi$, $\eta_c$ 
and $\chi_{c1}$ are calculated as 
$m_{J/\Psi} \simeq$ 3.07GeV, $m_{\eta_c} \simeq$ 2.99GeV and 
$m_{\chi_{c1}} \simeq$ 3.57GeV at $T\simeq 0$. 
These values almost the same as the experimental ones, 
$m_{J/\Psi}({\rm exp})= 3.10{\rm GeV}$,  
$m_{\eta_c}({\rm exp})=2.98{\rm GeV}$ and 
$m_{\chi_{c1}}({\rm exp})=3.51{\rm GeV}$.

\section{Temporal correlators of $c\bar c$ systems at finite temperature on
 anisotropic lattice} 
\label{sec4}

To investigate the low-lying state at high temperature 
from the temporal correlator, 
it is practically desired to use a 
``good" operator with a large overlap with ground state 
(ground-state overlap), due to limitation of the temporal lattice size. 
To this end, we use a spatially-extended operator 
of the Gaussian type as
\begin{eqnarray}
O(t,\vec x)\equiv N \sum_{\vec y}\exp\left\{-\frac{|\vec y|^2}{2\rho^2}\right\}
\bar c(t,\vec x+\vec y)\Gamma c(t,\vec x)
\end{eqnarray}
with the extension radius $\rho$ as the hadronic size 
in the Coulomb gauge \cite{IDIOOS05,MOU01}. 
$N$ is a normalization constant.
$\Gamma=\gamma_k (k=1,2,3)$ and $\Gamma=\gamma_5$ 
correspond to  $1^- (J/\Psi)$ and $0^- (\eta_c)$ channels, respectively. 
Note here that this form (14) is suitable for S-wave states (see Fig.~3(a)). 
In the actual calculation, 
the size parameter $\rho$ is optimally chosen in terms of 
the ground-state overlap. 
The energy of the low-lying state is calculated from 
the temporal correlator, 
\begin{eqnarray}
G(t)\equiv \frac{1}{V}\sum_{\vec x}\langle O(t,\vec x)
O^\dagger (0,\vec 0)\rangle,
\end{eqnarray}
where the total spatial 
momentum of the $c\bar c$ system is projected to be zero.

In accordance with the temporal periodicity at finite temperature, 
we define the effective mass $m_{\rm eff}(t)$ 
from the correlator $G(t)$ by the cosh-type function as \cite{ISM02}
\begin{align}
\frac{G(t)}{G(t+1)}=\frac{\cosh [m_{\rm eff}(t)(t-N_t/2)]} 
{\cosh [m_{\rm eff} (t) (t+1-N_t/2)]}
\label{cosh}
\end{align}
with the temporal lattice size $N_t$. 
If the correlator $G(t)$ is saturated by the lowest level,  
it behaves as
\begin{align}
G(t) \simeq A\cosh \left[m_0(t-\frac{N_t}{2})\right],
\label{cosh2}
\end{align}
where $m_0$ is the energy of the lowest state. 
In this case, 
the solution $m_{\rm eff}(t)$ in Eq.~(16) coincides with 
the lowest-state energy $m_0$ in Eq.~(17).

To find the optimal value of the 
extension radius $\rho$, we set two criterions. First, an effective mass 
$m_{\rm eff}(t)$ of a temporal correlator $G(t)$ with $\rho$ 
should have plateau region in terms of $t$. 
Second, the ground-state component should be large enough in the 
plateau region of $m_{\rm eff}(t)$. 
For the estimation of the ground state overlap, 
we define $g_{\rm fit}(t)$ as \cite{ISM02}
\begin{eqnarray}
g_{\rm fit}(t)\equiv A^\prime \cosh[m^\prime (t-N_t/2)],
\end{eqnarray} 
where $A^\prime$ and $m^\prime$ are determined by 
fitting $g_{\rm fit}(t)$ to $G(t)/G(0)$ in the plateau region. 
In general, $g_{\rm fit}(0)\le 1$ is satisfied, and $g_{\rm fit}(0)=1$ is 
achieved if and only if $G(t)$ is completely dominated by the ground-state 
contribution in the whole region of $0\le t\le 1/T$. 
Therefore, we use $g_{\rm fit}(0)$ as an index of the ground-state overlap. 
Namely, if $g_{\rm fit}(0)$ with $\rho_1$ is larger than that with $\rho_2$, 
we regard the operator with $\rho_1$ as a better operator 
which has larger ground-state overlap 
 than that with $\rho_2$. 
Under these criterions, 
we determine the optimal operator in the range of 
$\rho=0 \ (0.1{\rm fm}) \ 0.5{\rm fm}$. 
For $\rho=0$ and 
$0.1{\rm fm}$, the effective masses are found to 
have no plateau region. Accordingly, 
we abandon these extension radii. 
Next, we examine $g_{\rm fit}(0)$ for $\rho=(0.2-0.5){\rm fm}$, and 
find that the optimal size of $\rho$ is $\rho=(0.2-0.3){\rm fm}$ at all 
temperatures for $c\bar c$ systems. Actually, the results of 
$\rho=0.2{\rm fm}$ and $0.3{\rm fm}$ are almost the same. 
Hereafter, we show the numerical results for $\rho=0.2{\rm fm}$. 

\section{Lattice QCD results for $J/\Psi$ and $\eta_c$ channels above $T_c$}
\label{sec5}

We investigate the $c\bar c$ systems above $T_c$ 
both in $J/\Psi (J^P=1^-)$ and $\eta_c (J^P=0^-)$ channels in lattice QCD.
For each channel, 
we calculate the temporal correlator $G(t)$ 
following Eq.~(15), where the total spatial momentum is projected to be zero. 
From $G(t)$, we extract the effective mass $m_{\rm eff}(t)$ both on 
 PBC and APBC, 
and examine their spatial boundary-condition dependence. 
 From the cosh-type fit for the correlator $G(t)$ 
in the plateau region of effective mass $m_{\rm eff}(t)$, 
we extract the energies, $E$(PBC) and $E$(APBC), 
of the low-lying $c\bar c$ system on PBC and APBC, respectively. 
In fact, if the $c\bar c$ system is a spatially compact bound state, 
one can expect $E({\rm PBC})\simeq E({\rm APBC})$, which coincides with 
the bound state mass. 
On the other hand, if the $c\bar c$ system is a scattering state, 
$E$(PBC) and $E$(APBC) are 
expected to be the corresponding thresholds
 of the $c\bar c$ scattering state, i.e., 
$E({\rm PBC})\simeq 2m_c$ and $E({\rm APBC})\simeq 2\sqrt{m_c^2+
\vec{p}_{\rm min}^2}$. 

Figure 4 shows effective-mass plots of 
$J/\Psi$ on PBC and APBC for $(1.11-2.07)T_c$ in lattice QCD. 
Table \ref{table5} 
shows the boundary-condition dependence of 
the energy of the $c\bar c$ system
 in the $J/\Psi$ channel for $(1.11-2.07)T_c$. 
It is remarkable that almost no spatial boundary-condition dependence 
is found for the low-lying energy of the $c \bar c$ system, 
i.e., $\Delta E \equiv E{\rm (APBC)}-E{\rm (PBC)} \simeq 0$.
This contrasts to the $c \bar c$ scattering case of 
$\Delta E \simeq 2\sqrt{m_c^2+3\pi^2/L^2}-2m_c\simeq $ 0.35GeV 
for $L \simeq$1.55fm and $m_c \simeq$ 1.3GeV as was discussed 
in Sec.~\ref{sec2}.
This result indicates that $J/\Psi$ survives for $(1.11-2.07)T_c$ 
as spatially compact (quasi-)bound state. 
Figure 5 shows the temperature dependence of the pole mass 
of $J/\Psi$, $M_{J/\Psi}(T)$. 

Figure 6 shows effective mass plots of $\eta_c$ 
on PBC and APBC for $(1.11-2.07)T_c$. 
Table \ref{table6} summarizes the $c \bar c$ system 
in the $\eta_c$ channel on PBC and APBC at each temperature.
Again, almost no spatial boundary-condition dependence is found as 
$\Delta E \equiv E{\rm (APBC)}-E{\rm (PBC)}\simeq 0$. 
This result indicates that $\eta_c$ also survives 
for $(1.11-2.07)T_c$ as spatially compact 
(quasi-)bound state. 
Figure 7 shows the temperature dependence of the pole mass 
of $\eta_c$, $M_{\eta_c}(T)$. 


From Figs.~4 and 6,
one finds that $m_{\rm eff}(t)$ on APBC is saturated by 
the low-lying state more rapidly than that on PBC, 
i.e., $m_{\rm eff}^{\rm PBC}(t)>m_{\rm eff}^{\rm APBC}(t)$ for small $t$ 
and $m_{\rm eff}^{\rm PBC}(t)\simeq m_{\rm eff}^{\rm APBC}(t)$ for large $t$. 
This fact may be explained as follows. 
 In the channels of $J/\Psi$ and $\eta_c$, 
 there are scattering states in addition to the bound state.
On APBC, the low-lying scattering states are shifted to higher energy region 
in contrast to the compact bound state. 
Thus, the contributions of scattering states 
dump rapidly on APBC. 
The (slight) smallness of errorbars on APBC compared with those on PBC 
may be explained by the same reason (see Figs.~5 and 7). 

In Fig.~8, we compare $M_{J/\Psi}(T)$ with $M_{\eta_c}(T)$ on PBC.  
 It is interesting that there occurs 
 the inversion of $M_{J/\Psi}(T)$ and  $M_{\eta_c}(T)$ above $1.3T_c$. 
In fact, $M_{J/\psi}(T)$ decreases as temperature increases, while 
$M_{\eta_c}(T)$ is almost unchanged (see Fig.~5 and 7).  
We observe the significant reduction of $M_{J/\Psi}(T)$ of the 
order of 100MeV. 
(We note that thermal width broadening also leads to 
the same effect of the pole-mass reduction \cite{ISM02}.) 
In any case, it would be interesting to investigate the possible change 
of the $J/\Psi$ mass above $T_c$ . 

\begin{figure}[t]
\rotatebox{-90}{\includegraphics[width=5.2cm]
{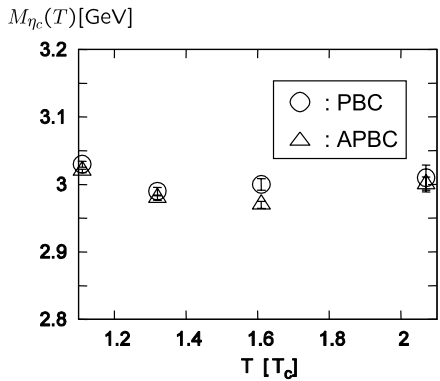}}

\caption{Temperature dependence of the pole mess (energy) $M_{\eta_c}(T)$ 
of $\eta_c$ for $(1.11-2.07)T_c$ on PBC (circles) and APBC (triangles). 
The energy difference $\Delta E\simeq 0$ 
between PBC and APBC are also considerably 
smaller than that of $c\bar c$ scattering state as well as $J/\Psi$ 
case.}
\label{fig11}
\end{figure}
\begin{figure}[t]
\rotatebox{-90}{\includegraphics[width=5.2cm]
{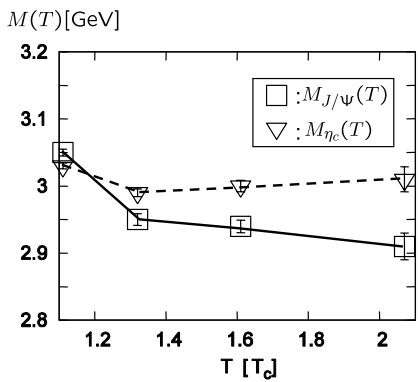}}
\caption{Temperature dependence of the pole mass (on PBC) 
of $J/\Psi$ and $\eta_c$ for $(1.11-2.07)T_c$. 
The squares denote $M_{J/\Psi}(T)$ and 
the inverse triangles denote $M_{\eta_c}(T)$. 
There occurs the level inversion of $J/\Psi$ and $\eta_c$ above $1.3T_c$.}
\label{fig12}
\end{figure}

\section{MEM analysis for $J/\Psi$, $\eta_c$ and $\chi_{c1}$ channels above $T_c$}
\label{sec6}
In this section, we perform the MEM analysis for the $c\bar c$ systems in 
$J/\Psi$ ($J^P=1^-$), $\eta_c$ ($J^P=0^-$) and $\chi_{c1}$ ($J^P=1^+$) channels 
above $T_c$ using the lattice QCD data.
Here, the axial vector $\chi_{c1}$ is a $P$-wave $c\bar c$ meson and 
its wave function tends to spread due to the centrifugal potential 
$l(l+1)/m_c r^2$ ($r$: relative distance between $c$ and $\bar c$, 
$l$: orbital angular momentum) 
compared with $S$-wave states such as $J/\Psi$ and $\eta_c$ (see Fig.~3). 
According to this spread wave function, $\chi_{c1}$ is expected to be 
sensitive to vanishing of the linear potential and appearance of the 
Debye screening effect above $T_c$. 
In fact, the dissociation temperature of $\chi_{c1}$ would be 
lower than that of $J/\Psi$ and $\eta_c$. 
Therefore, we study the $P$-wave $c\bar c$ meson $\chi_{c1}$ 
at finite temperature in lattice QCD. 
%

Here, we note that the difficulty with the extended operator in 
axial vector channel.  
Because $\chi_{c1}$ is a $P$-wave state, 
the wave function of $\chi_{c1}$ is not spherical. 
Moreover, its radial wave function should be zero at the origin due to 
the centrifugal potential, as shown in Fig.~3(b). 
Therefore, the spherical and Gaussian type extension of the operator, 
which is suitable 
for an $S$-wave state (see Fig.~3(a)), 
 may not be a good choice to investigate a 
possible $P$-wave bound state of $\chi_{c1}$. 
Actually, we calculate the effective mass with Gaussian extended 
operator with $\rho=(0-0.5){\rm fm}$ in $\chi_{c1}$ channel above $T_c$ and 
find that the effective mass has no plateau region. 
This fact indicates that we cannot extract the low-lying energy state clearly 
in $\chi_{c1}$ channel with the operator. 


Instead, we perform the analysis of $\chi_{c1}$ above $T_c$ with maximally entropy method (MEM)
 \cite{AH04,UNM02,DKPW04} with local interpolating field, $O_{\chi_{c1}}=\bar c \gamma_5\gamma_k c$.
Using MEM, we can extract the spectral function $A(\omega)$ 
from the temporal correlator $G(t)$, 
where the relation between $A(\omega)$ and $G(t)$ at finite temperature 
$T$ is given by 
\begin{eqnarray}
G(t)&=&\int_0^\infty d\omega K(t,\omega)A(\omega), \nonumber\\
K(t,\omega)&=& (e^{-t \omega}+e^{-(1/T-t)\omega})/(1-e^{-\omega/T}). 
\end{eqnarray}
We use the Wilson quark action in the calculation of lattice QCD data 
for MEM. Instead of the improvement of the action, 
we use the finer lattice spacing 
$a_t=a_s/4=(20.2{\rm GeV})^{-1}=9.75\times 10^{-3}{\rm fm}$ with $\beta=7.0$. 
The lattice size is $20^3\times 46$, which corresponds to the 
lattice volume $L^3\simeq (0.78{\rm fm})^3$ and the temperature 
$T=1.62T_c$. 

To begin with, we show the MEM results for $J/\Psi$ and $\eta_c$. 
Figures 9 and 10 show the spectral function of $J/\Psi$ and $\eta_c$, respectively, with PBC and APBC. 
Clear low-lying peaks corresponding to $J/\Psi$ and $\eta_c$ are observed around $3{\rm GeV}$. 
Here, we note that the appearance of 
the peak structure is highly nontrivial, because 
the default function of MEM is 
a perturbative one, which does not have the peak structure. 
No difference between PBC and APBC is observed for $J/\Psi$ and $\eta_c$, 
which indicates that $J/\Psi$ and $\eta_c$ appear as spatially-localized compact bound states. 
These results of $J/\Psi$ and $\eta_c$ are consistent with those obtained in previous sections. 
The peaks in high-energy region ($\omega >5{\rm GeV}$) are considered as lattice artifacts  
\cite{DKPW04,CPPACS-Yamazaki}. 

Now, we investigate the MEM analysis for $\chi_{c1}$ at finite temperature in lattice QCD. 
Figure 11(a) shows the spectral function in $\chi_{c1} (J^P=1^+)$ channel 
on PBC at $1.62T_c$. 
There is no low-lying peak which corresponds to $\chi_{c1}$ 
($m_{\chi_{c1}}\simeq 3.5{\rm GeV}$). 
In fact, in contrast to $J/\Psi$ and $\eta_c$, 
the low-lying structure of the spectral function in $\chi_{c1}$ channel differs from that in 
$J/\Psi$ and $\eta_c$ channels at $1.62T_c$.
Therefore, the MEM analysis indicates that the dissociation of 
$\chi_{c1}$ occurs already at $1.62T_c$. 
In the high-energy region around $6{\rm GeV}$, we can see a sharp peak 
in the spectral function. 
Figure 11(b) shows the spectral function in $\chi_{c1}$ channel on APBC 
at $1.62T_c$. 
There is almost no difference between PBC and APBC. 
We compare these results in Fig.~11(c) and confirm that the spectral function 
on PBC almost coincides with that on APBC. 
This BC-independence indicates that the peak around $6{\rm GeV}$ corresponds to
 a spatially compact (quasi-)bound state. 
 The state around 6GeV may be a bound state of doubler(s), as was suggested in Refs.~\cite{DKPW04,CPPACS-Yamazaki}. 
\begin{figure*}[htb]
\includegraphics[width=6.5cm]
{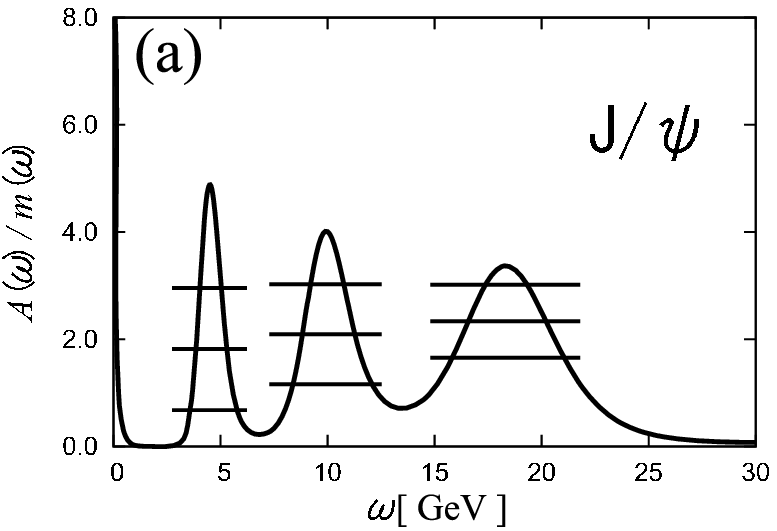}
\includegraphics[width=6.5cm]
{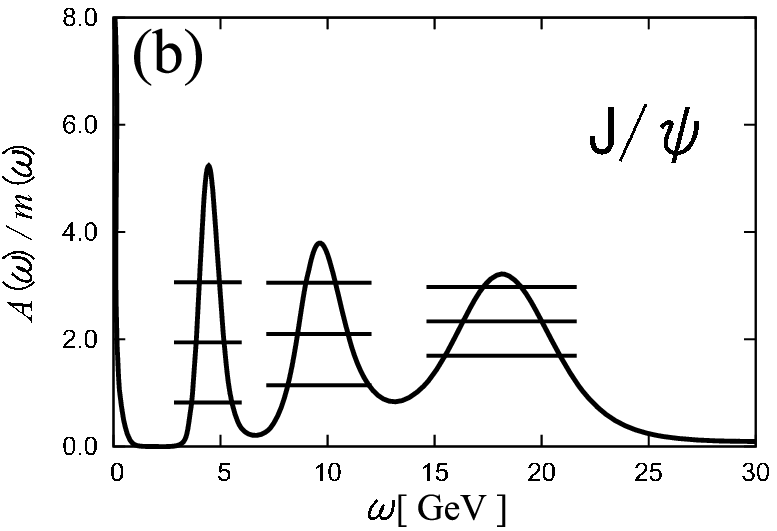}
\includegraphics[width=6.5cm]
{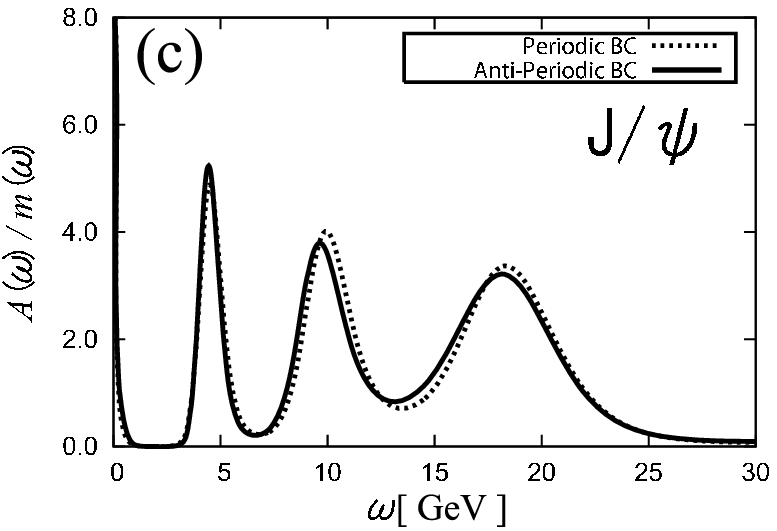}
\caption{The spectral function $A(\omega)$ 
of the $c\bar c$ state in vector ($J^P=1^-$) 
channel on PBC (a) and APBC (b)
at $1.62T_c$ extracted 
with MEM from lattice QCD data of the temporal correlator. 
$m(\omega)$ denotes the default model of MEM \cite{AH04,UNM02,DKPW04}. 
 The statistical error is denoted by the three solid lines. 
There is a low-lying peak which corresponds to $J/\Psi \ (m_{J/\Psi}\simeq 3.1{\rm GeV})$ 
even above $T_c$.
We add (c) as the comparison between PBC and APBC. These results almost coincide.}
\label{MEM-jpsi}
\end{figure*}

\begin{figure*}[htb]
\includegraphics[width=6.5cm]
{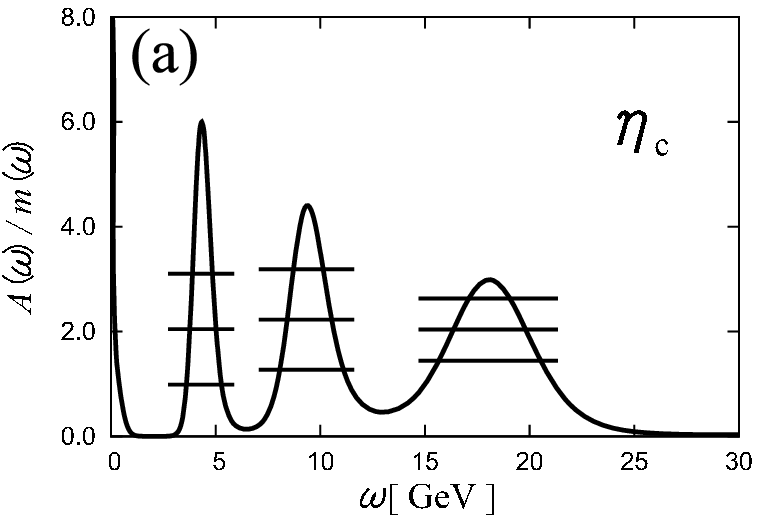}
\includegraphics[width=6.5cm]
{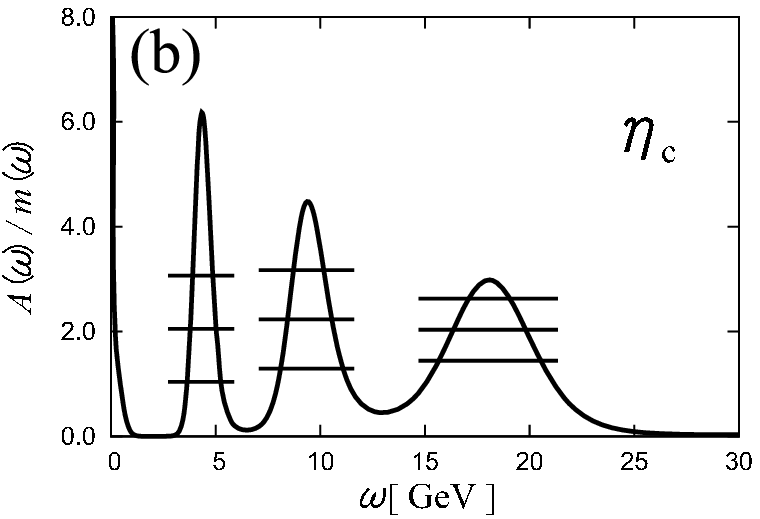}
\includegraphics[width=6.5cm]
{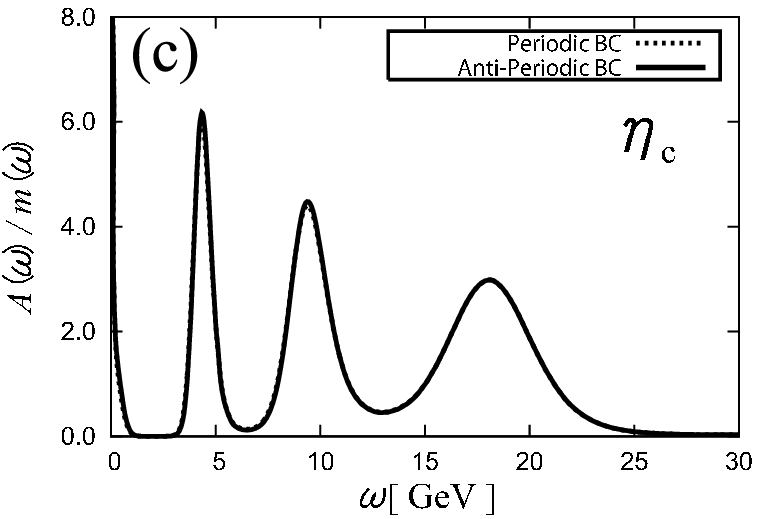}
\caption{The spectral function $A(\omega)$ 
of the $c\bar c$ state in pseudo scalar ($J^P=0^-$) 
channel. 
The notations are the same as those in the $J/\Psi$ case (Fig.~9). 
There is a low-lying peak which corresponds to $\eta_c \ (m_{\eta_c}\simeq 3.0{\rm GeV})$. 
Because the results in PBC and APBC cases almost completely coincide,
 the dotted line is blinded by the solid line in Fig.~10(c).}
\label{MEM-chic1}
\end{figure*}

\begin{figure*}[htb]
\includegraphics[width=6.5cm]
{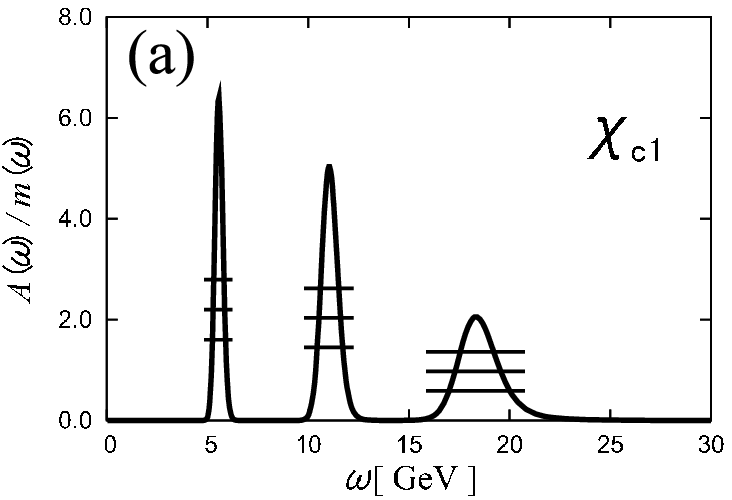}
\includegraphics[width=6.5cm]
{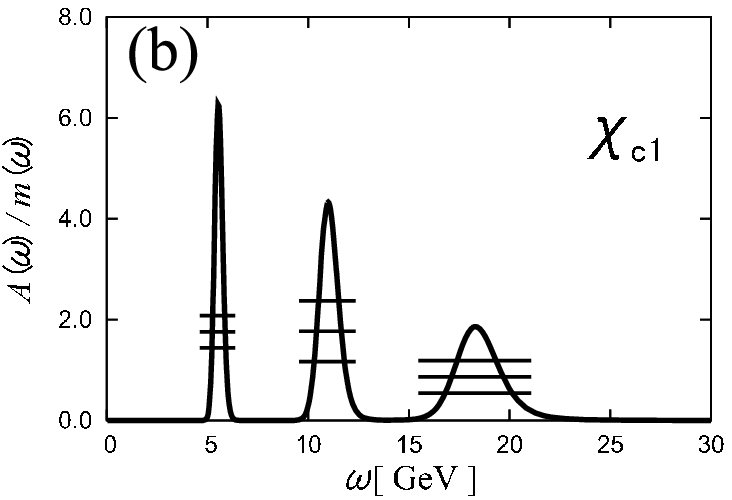}
\includegraphics[width=6.5cm]
{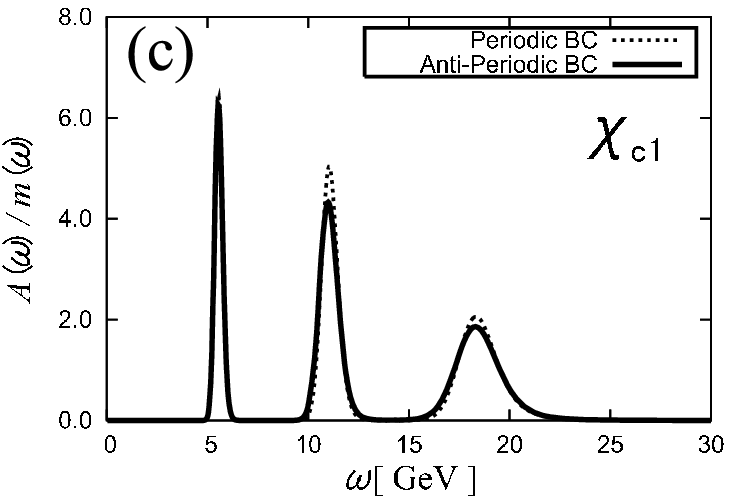}
\caption{The spectral function $A(\omega)$ 
of the $c\bar c$ state in axial vector ($J^P=1^+$) 
channel on PBC (a) and APBC (b)
at $1.62T_c$ extracted 
with MEM from lattice QCD data of the temporal correlator. 
$m(\omega)$ denotes the default model of MEM \cite{AH04,UNM02,DKPW04}. 
 The statistical error is denoted by the three solid lines. 
In contrast to the $J/\Psi$ and $\eta_c$ cases, there is no low-lying peak which corresponds to $\chi_{c1} 
(m_{\chi_{c1}}\simeq 3.5{\rm GeV})$. 
The peaks in high-energy region ($\omega >5{\rm GeV}$) are considered as 
 lattice artifacts. 
We add (c) as the comparison between PBC and APBC. These results almost 
coincide.}
\label{MEM-chic1}
\end{figure*}




\section{Conclusion and outlook}
\label{sec7}

We have investigated $J/\Psi$ and $\eta_c$ above $T_c$ with anisotropic 
quenched lattice QCD to clarify whether the $c\bar c$ systems above $T_c$ are 
compact (quasi-)bound states or scattering states.
We have adopted the standard Wilson gauge action and 
the $O(a)$-improved Wilson quark action with renormalized anisotropy 
$a_s/a_t=4.0$. 
We have used $\beta=6.10$ on $16^3\times (14-26)$ lattices, 
which correspond to the spatial lattice volume 
$V\equiv L^3\simeq (1.55{\rm fm})^3$ and $T=(1.11-2.07)T_c$. 
To clarify whether compact charmonia survive in the deconfinement phase, 
we have investigated spatial boundary-condition dependence of the energy 
of $c\bar c$ systems above $T_c$. 
In fact, for low-lying $c\bar c$ scattering states, 
it is expected that there appears a significant energy difference $\Delta E
\equiv E({\rm APBC})-E({\rm PBC})$ 
between periodic and anti-periodic boundary 
conditions as $\Delta E\simeq 2\sqrt{m_c^2+3\pi^2/L^2}-2m_c$ on the 
finite-volume lattice. 
For enhancement of the ground-state overlap, we have used the spatially 
extended operator with Gaussian function. 

As a result, both in $J/\Psi$ and $\eta_c$ channels, 
we have found almost no spatial boundary-condition dependence of 
the energy of the low-lying $c\bar c$ system 
 even on the finite-volume lattice for $(1.11-2.07)T_c$. 
These results indicate that $J/\Psi$ and $\eta_c$ 
survive as spatially compact $c \bar c$ (quasi-) bound states 
for $(1.11-2.07)T_c$. 
Also, the inversion of levels of $J/\Psi$ and $\eta_c$ above 
$1.3T_c$ has been seen. 
In fact, we have observed the 
significant reduction of the $J/\Psi$ pole mass of about $100{\rm MeV}$, 
above $1.3T_c$. 
Experimentally, it may be interesting to investigate the possible change 
of the $J/\Psi$ mass above $T_c$.

We have also performed the MEM analysis for the $c\bar c$ systems in 
$J/\Psi$ ($J^P=1^-$), $\eta_{c}$ ($J^P=0^-$) and 
$\chi_{c1}$ ($J^P=1^+$) channels above $T_c$ using the lattice QCD data. 
For this analysis, we have adopted lattice QCD 
at $1.62T_c$ with the Wilson quark action with 
$\beta=7.0$ and $a_s/a_t=4.0$. 
For the $S$-wave channel, we have obtained  
the same results on the survival of $J/\Psi$ and $\eta_c$ as the spatially-localized 
compact bound state above $T_c$. 
In contrast to the $J/\Psi$ and $\eta_c$ cases, 
 the spectral function in the $P$-wave channel 
has no low-lying peak structure corresponding to $\chi_{c1}$ around $3.5{\rm GeV}$. 
This fact indicates that $\chi_{c1}$ already dissociates at $1.62T_c$. 
 The further analysis of charmonia at high temperature 
 with MEM will be reported in Ref.~\cite{tumura}.

Through the MEM analysis, 
we have observed the spatially localized bound state in high energy region, 
which is not affected by the spatial boundary condition at all. 
The bound state appearing 
in high-energy region on the lattice would be the bound state of 
doubler(s), which is unphysical. 

Our study of charmonia indicates that the $S$-wave mesons 
$J/\Psi$ and $\eta_c$ survive 
even above $T_c (\sim 2T_c)$ as compact bound states.

As a successive work, we are performing 
the same analyses for other charmed mesons, e.g., 
$D$ mesons.
 The narrowness 
of the decay width of charmonia strongly depends on 
whether the decay channel of $D\bar D$ opens or not. 
If the mass of the $D$ meson is shifted
 at high temperature, a drastic change of decay width of 
some charmonia possibly occurs, which may be 
interesting both theoretically and experimentally \cite{H00}. 

The subject of hadrons in high density system is fascinating. 
For example, CERES collaboration presented interesting events 
on the lepton pair production in low-mass region from 
p-Be and p-Au ($450{\rm GeV}/c$) 
and from S-Au ($200{\rm GeV}/c$ per nucleon) \cite{CERES95}.
It is reported that the lepton pairs from S-Au in the invariant mass range 
$0.2{\rm GeV}/c^2<m<1.5{\rm GeV}/c^2$ 
are largely enhanced compared with the hadronic contributions, 
while those from p-Be and p-Au are not enhanced. 
The enhancement is considered to be the signal of 
the change of hadronic nature of 
vector mesons in finite density system \cite{CERES95, EIK98, IOS05B}. 
Namely, the mass reduction and/or the width broadenings of vector mesons
at finite density can lead to the enhancement of low-energy lepton pairs. 
At the experimental facility J-PARC in Japan, 
which is planned to run in 2008, the experiments of high density system 
are expected to take place. 
For these experiments, 
the theoretical study of finite density system is required. 
Then the study at finite density in lattice QCD 
is challenging and worth trying. 

 The survival of $J/\Psi$ above $T_c$ 
may change the scenario of $J/\Psi$ suppression. 
 These analyses give us the further knowledge of QCD at finite 
 temperature.

\begin{acknowledgements}
{We thank Prof.~M.~Oka for fruitful discussions 
and suggestions. 
H.~S. and N.~I. are supported in part by the Grant for Scientific Research 
[(B) No. 15340072 and (C) No. 16540236] from the 
Ministry of Education, Culture, Sports, Science and Technology, Japan. 
H.~I. is supported by the Japan Society for the Promotion of Science for 
Young Scientists. 
T.~D. is supported by 
Special Postdoctoral Research Program of RIKEN. 
K.~T. is supported by a 21st Century COE Program at Kyoto University. 
Our lattice QCD calculations have been performed on 
NEC-SX5 at Osaka University.}
\end{acknowledgements}

\end{document}